\documentclass[11pt,a4paper]{article}
\usepackage{graphicx}
\usepackage{float}
\usepackage{afterpage}
\usepackage{epsfig,cite}
\usepackage{amssymb}
\usepackage{amsmath}
\usepackage{dsfont}
\usepackage{multirow}
\usepackage{url,hyperref}
\usepackage{booktabs}
\usepackage{tabularx}
\usepackage{xcolor}
\usepackage{bm}
\usepackage{textcomp}
\usepackage{caption}
\usepackage{slashed}
\usepackage{changes}
\usepackage{url}
\usepackage{hyperref}

\usepackage[normalem]{ulem}

\usepackage{url}
\usepackage{hyperref}

\begin{document}

\vspace{-2.0cm}

\begin{center} 

  {\large \bf 
Matching generalised transverse-momentum-dependent distributions\\
\vspace{5pt}
  onto generalised parton distributions at one loop}
  \vspace{.7cm}

Valerio~Bertone

\vspace{.3cm}
{\it IRFU, CEA, Universit\'e Paris-Saclay, F-91191 Gif-sur-Yvette, France}

\end{center}   

\vspace{0.1cm}

\begin{center}
  {\bf \large Abstract}\\
\end{center}

The operator definition of generalised transverse-momentum-dependent
(GTMD) distributions is exploited to compute for the first time the
full set of one-loop corrections to the off-forward matching
functions. These functions allow one to obtain GTMDs in the
perturbative regime in terms of generalised parton distributions
(GPDs). In the unpolarised case, non-perturbative corrections can be
incorporated using recent determinations of
transverse-momentum-dependent (TMD) distributions. Evolution effects
for GTMDs closely follow those for TMDs and can thus be easily
accounted for up to next-to-next-to-leading logarithmic accuracy. As a
by-product, the relevant one-loop anomalous dimensions are derived,
confirming previous results. As a practical application, numerical
results for a specific kind of GTMDs are presented, highlighting some
salient features.

\newpage

\tableofcontents

\section{Introduction}

At present, the study of the hadronic structure is a particularly
lively field of research. If on the one hand this is a very
interesting topic in itself, on the other hand it is instrumental to
precision physics at present and future high-energy colliders.

The case of the unpolarised collinear parton-distribution functions
(PDFs) of the proton is emblematic of the huge effort that is being
put into the study of the hadronic structure. Driven by the
experimental activity of colliders such as HERA, Tevatron, and the
Large Hadron Collider (LHC), and by a steady methodological and
theoretical progress, PDFs are nowadays known with an astonishing
precision~\cite{PDF4LHCWorkingGroup:2022cjn, Hou:2019efy,
  Bailey:2020ooq, NNPDF:2021njg, McGowan:2022nag}. Driven by the need
for accuracy, also the study of unpolarised collinear fragmentation
functions (FFs) has recently been intensified leading to accurate
determinations~\cite{Sato:2019yez, Abdolmaleki:2021yjf,
  Moffat:2021dji, Khalek:2021gxf, Borsa:2022vvp, Khalek:2022vgy}.

Although PDFs and FFs have a broad phenomenological applicability,
they encode partial information on the hadronic structure
corresponding to the longitudinal-momentum distribution of partons
inside hadrons. Transverse-momentum-dependent (TMD) distributions are
instead also sensitive to the transverse momentum of partons thus
extending the information provided by PDFs and
FFs~\cite{Collins:1984kg, Collins:2011zzd}. Also due to their
relevance in hot topics such as the precision determination of the
mass of the $W$ boson, TMDs are currently receiving particular
attention and much work is being invested in their
determination~\cite{Bacchetta:2017gcc, Bacchetta:2019sam,
  Scimemi:2019cmh, Bacchetta:2022awv, Boglione:2022nzq,
  Ebert:2022cku}.

Another typology of distributions relevant to the study of the
hadronic structure is that of generalised parton distributions
(GPDs). These distributions give us access to the energy-momentum
tensor of hadrons~\cite{Ji:1998pc,Polyakov:2002yz}, providing us with
a handle on important quantities like the transverse position of
partons~\cite{Burkardt:2002hr,Diehl:2002he} and their angular
momentum~\cite{Ji:1996ek}. Phenomenological determinations of GPDs do
exist~\cite{Kumericki:2015lhb,Dutrieux:2021wll, Guo:2022upw} but, as
of today, they are much less developed than modern analyses of
PDFs/FFs and TMDs. However, the recent approval of the electron-ion
colliders in China (EicC)~\cite{Anderle:2021wcy} and in the US
(EIC)~\cite{AbdulKhalek:2021gbh}, has revived the interest in GPDs
that is now a rapidly growing field. In addition, new technologies on
lattice~\cite{Ji:2013dva, Radyushkin:2017cyf} have made
first-principle computations of GPDs more accessible, giving
additional momentum to their study.

It turns out that PDFs, TMDs, and GPDs are all projections of more
general quantities, usually referred to as generalised TMDs (GTMDs),
that can be considered as their mother
distributions~\cite{Belitsky:2003nz, Ji:2003ak, Meissner:2008ay,
  Meissner:2009ww, Lorce:2011kd, Lorce:2011ni, Lorce:2013pza}. As of
today, very little is quantitatively known about GTMDs and most of the
existing studies are based on models~\cite{Lorce:2011dv,
  Kanazawa:2014nha, Lorce:2015sqe, Hagiwara:2016kam, Hagiwara:2017fye,
  More:2017zqq, More:2017zqp, Bhattacharya:2018zxi,
  Mantysaari:2019csc, Kaur:2019kpi, Salazar:2019ncp, Luo:2020yqj,
  Zhang:2020ecj, Boer:2021upt}. In spite of recent proposals to access
GTMDs experimentally~\cite{Hatta:2016dxp, Bhattacharya:2017bvs,
  Bhattacharya:2018lgm}, data sensitive to these distributions is
presently scarce, making phenomenological studies laborious.

The goal of this paper is to exploit as much as possible our knowledge
of the projections of GTMDs, namely GPDs and TMDs, to reconstruct
GTMDs themselves to the best accuracy possible. To this purpose, a
proper definition of GTMDs has to be devised, which enables the
computation of relevant perturbative quantities. In the spirit of TMD
factorisation, this was done in Ref.~\cite{Echevarria:2016mrc} where,
generalising the TMD operators to the off-forward case, a
rapidity-divergence-free definition of GTMD correlators was given. We
will start from this definition to compute for the first time the full
set of the so-called matching functions at one-loop accuracy. These
functions allow one to obtain GTMDs in terms of GPDs by accounting for
the emission of partons with large transverse momentum $\mathbf{k}_T$,
whose effect is thus computable in perturbation theory. We will
finally use these matching functions to reconstruct realistic GTMDs
also accounting for evolution and non-perturbative effects.

The outline of the paper is as follows. In
Sect.~\ref{sec:operator_definition}, we will give a precise operator
definition of the GTMD correlators of our interest.
Sect.~\ref{sec:renormalisation_and_evolution} is devoted to the
renormalisation of the UV divergences of these correlators and to the
derivation of their evolution equations. In Sect.~\ref{sec:matching}
we will state the GTMD matching formula that factorises
large-$\mathbf{k}_T$ emissions into a set of matching functions to be
convoluted with GPDs. In this section, we will exploit this matching
formula to express the one-loop matching functions in terms of
perturbative quantities, namely the soft function and the unsubtracted
parton-in-parton GTMD correlator, whose one-loop corrections are
computed in Sects.~\ref{sec:SoftFunction}
and~\ref{sec:unsubtractedGTMDs}, respectively. At this point, we will
be in a position to obtain the explicit expression for the one-loop
matching functions. In Sect.~\ref{sec:forward_limit}, we will show
that, as expected, these expressions tend to their TMD counterpart in
the forward limit. As a by-product of UV renormalisation, in
Sect.~\ref{sec:anomalous_dimensions} we will extract the anomalous
dimensions that govern the GTMD evolution at one loop, finding that
they agree with the TMD ones. A numerical implementation of the
matching functions, combined with other existing perturbative and
non-perturbative ingredients, puts us in a position to reconstruct
realistic quark and gluon GTMDs to high accuracy. A study of the
resulting distributions is presented in Sect.~\ref{sec:numerics} where
we will consider the behavior of GTMDs under different points of view,
commenting on some peculiar features. Finally, in
Sect.~\ref{sec:conclusions} we will draw our conclusions.

\section{Theoretical setup and results}

As argued in the pioneering Ref.~\cite{Echevarria:2016mrc}, a sound
definition of GTMDs, \textit{i.e.} a definition free of rapidity
divergences and that can thus be used for phenomenology, requires
taking into proper account the soft function. In this section, we will
review the reasoning of Ref.~\cite{Echevarria:2016mrc} that leads to a
rapidity-divergence-free definition of the GTMD correlator. Working in
the space of the Fourier-conjugate variable of the partonic momentum
$\mathbf{k}_T$, denoted by $\mathbf{b}_T$, we will formulate the
explicit definition of the GTMD correlator both for quarks and
gluons. We will then state the matching formula that, for
$\mathbf{b}_T\simeq \mathbf{0}$, allows us to express the GTMD
correlator in terms of (collinear) GPD correlators by means of
perturbatively computable matching functions. Appealing to the concept
of parton-in-parton distributions (see \textit{e.g.}
Ref.~\cite{Collins:2011zzd, Bertone:2022frx}), we will finally obtain
the one-loop corrections to these matching functions.

This programme requires the computation of the first perturbative
correction to the soft function and the so-called
\textit{unsubtracted} GTMD correlators. By combining these two
quantities, we will explicitly exhibit the cancellation of the
rapidity divergences and obtain, for the first time, the full set of
off-forward matching functions at one loop. In addition, we will show
that their forward limit coincides with the one-loop TMD matching
functions, as it should. Finally, as a by-product of the
renormalisation of the UV divergences, we will also derive the GTMD
evolution equations and extract the leading-order term of the relevant
anomalous dimensions, confirming previous results.

The seminal work of Refs.~\cite{Meissner:2009ww, Lorce:2013pza} has
provided us with a thorough classification of the structures that
emerge from the analysis of the GTMD correlators for spin-1/2
targets. However, if taken literally, the GTMD definitions given in
those papers produce divergent results upon inclusion of radiative
corrections, even after the customary renormalisation of the UV
divergences. Exactly like in the TMD case~\cite{Collins:2011zzd,
  Echevarria:2011epo}, these spurious divergences are of IR origin and
stem from the region of loop momenta $k$ where the rapidity of the
emitted partons, $y=\ln(k^+/k^-)$, becomes
large~\cite{Collins:2003fm}: hence they are usually called rapidity
divergences. This ``inconvenience'' can be fixed in two main steps:
\begin{enumerate}
\item removing the overlap of the GTMD correlator with the soft modes
  by means of the so-called zero-bin (or soft) subtraction,
\item redistributing the soft function, that typically appears in a
  factorisation theorem along with two beam functions, between the two
  GTMD correlators: this usually amounts to assigning the square root
  of the soft function to each of them.
\end{enumerate}
An implementation of these steps is guaranteed to lead to a
cancellation of the rapidity divergences. The reader is referred to,
\textit{e.g.}, Refs.~\cite{Echevarria:2016scs, Ebert:2019okf} and
references therein for a more detailed discussion in the TMD
framework.

In order to perform an explicit perturbative calculation, rapidity
divergences need to be regulated on a diagram-by-diagram basis.
Several rapidity-divergence regulators have been proposed so far in
the literature. However, for a specific class of
regulators~\cite{Idilbi:2010im, Ebert:2019okf}, steps 1) and 2)
combine in a way that the net effect is that the unsubtracted GTMD
correlator has to be \textit{divided} by the square root of the soft
function. In this paper, we will use the principal-value regulator,
originally introduced in Ref.~\cite{Curci:1980uw}, that indeed belongs
to this class of regulators.

\subsection{Operator definition of GTMDs}\label{sec:operator_definition}

This brief introduction allows us to finally give the exact definition
for the quark and gluon GTMD correlators that will be used for the
computation of the one-loop matching functions. As anticipated above,
it is convenient to work in $\mathbf{b}_T$ space which simply amounts
to taking the Fourier transform of the $\mathbf{k}_T$-space GTMD
correlators. In addition, for definiteness, we will limit the
discussion to the specific twist-2 GTMD correlators whose forward
limit gives the unpolarised quark and gluon TMDs. The corresponding
operator definition for a generic hadronic target $H$ is then:
\begin{equation}
\hat{\mathcal{F}}_{i/H}(x,\xi,\mathbf{b}_T,t)=\hat{S}_i^{-\frac12}(\mathbf{b}_T)
\hat{\Phi}_{i/H}(x,\xi,\mathbf{b}_T,t)\,,\quad i=q,g\,,
\label{eq:subGTMD}
\end{equation}
where $\hat{S}_i$ is the appropriate soft function and
$\hat{\Phi}_{i/H}$ is the unsubtracted GTMD correlator. The quark and
gluon correlators respectively read:
\begin{eqnarray}
\nonumber \displaystyle \hat{\Phi}_{q/H}(x,\xi,\mathbf{b}_T,t)
  &=&\displaystyle \int\frac{dy}{2\pi}e^{-ix(n\cdot P)y}\left\langle
  P_{\rm out}\left|\left[\overline{\psi}_q W_{n,q}^\dag\right]\left(\frac{\eta}2\right)\frac{\slashed{n}}2
  \left[W_{n,q}\psi_q\right]\left(-\frac{\eta}2\right)\right|P_{\rm in}\right\rangle\,,\\
\label{eq:GTMDdefinition}  \\
\nonumber  \displaystyle \hat{\Phi}_{g/H}(x,\xi,\mathbf{b}_T,t) &=&\displaystyle \frac{n_\mu n_\nu }{x(n\cdot P)}\int\frac{dy}{2\pi}e^{-ix(n\cdot P)y}\left\langle
  P_{\rm out}\left|\left[F_{a}^ {\mu j}W_{n,g}^\dag\right]\left(\frac{\eta}2\right)
  \left[W_{n,g} F_{a}^{\nu j}\right]\left(-\frac{\eta}2\right)\right|P_{\rm in}\right\rangle\,,
\end{eqnarray}
where we have used the following shorthand definitions:
$\eta = yn+\mathbf{b}_T$\footnote{Here, the transverse vector
  $\mathbf{b}_T$ is to be interpreted as a four-dimenesional embedding
  with null time and longitudinal components and transverse component
  coinciding with $\mathbf{b}_T$. This notation is repeatedly used
  below.}, $P_{\rm in/out}=P\pm\Delta/2$, $t=\Delta^2$,
$\xi=2n\cdot \Delta / n\cdot P$. Here, $\psi_q$ indicates the spinor
of the quark flavour $q$ and $F_a^{\mu\nu}$ the gluon field strength,
while $W_{n,i}$ is the Wilson line in the $n$ direction. The integrals
run between $-\infty$ and $+\infty$ and a summation over repeated
Lorentz indices is understood. Also the index $j$ in
$\hat{\Phi}_{g/H}$ is summed over and it runs over the (physical)
transverse components, \textit{i.e.} $j=1,2$. $n$ and $\overline{n}$
(the latter appearing below) are light-like four-vectors,
$n^2=\overline{n}^2=0$, and their scalar product evaluates to
$n\cdot \overline{n}=1$.\footnote{In light-cone coordinates defined as
  $a^\mu=(a^+,a^-,\mathbf{a}_T)$ with $a^{\pm}=(a_t\pm a_z)/\sqrt{2}$,
  the components of $n$ and $\overline{n}$ are usually chosen to be
  $n^\mu=(0,1,\mathbf{0})$ and $\overline{n}^\mu=(1,0,\mathbf{0})$.}
Also notice that the parton index $i$, labeling both the soft function
$\hat{S}_i$ and the Wilson line $W_{n,i}$, denotes the colour-group
representation. Specifically, for $i=q(g)$ the fundamental (adjoint)
representation of the SU(3) generators is to be used.  Finally, the
symbol $\hat{\phantom{F}}$ over $\hat{S}_i$ and $\hat{\Phi}_{i/H}$
indicates that these quantities are bare, \textit{i.e.}  they are UV
divergent in four dimensions.

The definition of the GTMD correlators is still incomplete because we
have not specified $\hat{S}_i$ and $W_i$. In fact, the precise form of
these quantities depends on the choice of the gauge. Two specific
gauges are usually considered in this context: the Lorentz gauge
($\partial\cdot A=0$) and the light-cone gauge ($n\cdot A=0$). When
considering $\mathbf{k}_T$-integrated quantities, which corresponds to
setting $\mathbf{b}_T=\mathbf{0}$, the light-cone gauge is often
preferred. The reason is that the Wilson lines reduce to unity with a
consequent reduction of the number of diagrams to be considered (see,
\textit{e.g.}, Refs.~\cite{Curci:1980uw, Bertone:2022frx}). This
simplification, however, comes at the price of a more complicated
gluon propagator. Conversely, when $\mathbf{k}_T$ is left
unintegrated, so that $\mathbf{b}_T\neq\mathbf{0}$, two types of
Wilson line, longitudinal and transverse, enter the
game~\cite{Ji:2002aa, Idilbi:2010im}. It turns out that in light-cone
gauge the longitudinal Wilson lines unitarise but the transverse ones
do not~\cite{Ji:2002aa}. The opposite happens in Lorentz gauge where
the transverse Wilson lines unitarise while the longitudinal ones do
not. As a consequence, using the light-cone gauge at
$\mathbf{b}_T\neq\mathbf{0}$ brings no advantage over the Lorentz
gauge in that the presence of the transverse Wilson lines invalidates
the argument of a smaller number of diagrams.\footnote{It should be
  pointed out that the choice $n\cdot A=0$ does not entirely fix the
  gauge. Therefore, one can use the remaining freedom to set to zero
  the transverse component of the gauge field at light-cone infinity
  in a way that the transverse Wilson line also
  unitarises~\cite{Rodini:2021zcs}. However, this additional gauge
  condition can only be enforced at either positive or negative (or a
  combination of the two) light-cone infinity, implying that the
  transverse Wilson line cannot be made unitarise simultaneously for
  all kinematics. I am indebted to S. Rodini for drawing my attention
  to this aspect.} Yet, in light-cone gauge the gluon propagator
remains more complicated than in Lorentz gauge.  In conclusion, in the
GTMD case (as well as in the TMD one) the Lorentz gauge seems to be a
preferable choice for perturbative calculations and is the one adopted
here.

The choice of the gauge finally allows us to write the explicit form
of the soft function:
\begin{equation}
  \hat{S}_i(\mathbf{b}_T) =
  \frac{1}{N_i}\mbox{Tr}_c\langle0| W_{\overline{n},i} (\mathbf{b}_T)
  W_{n,i}^\dag (\mathbf{b}_T) W_{n ,i} (\mathbf{0}) W_{\overline{n},i}^\dag
  (\mathbf{0})|0\rangle\,,\quad i=q,g\,,
  \label{eq:softfunctiondefLorentz}
\end{equation}
with $N_q=N_c=3$ and $N_g=N_c^2-1=8$ being the dimensions of the
fundamental and adjoint representations of the colour group,
respectively. A trace over the colour indices is indicated by
$\mbox{Tr}_c$. The Wilson line, also entering the definition of the
unsubtracted GTMD correlators in Eq.~(\ref{eq:GTMDdefinition}), is
given by:
\begin{equation}
  W_{v,i} (\mathbf{b}_T) = \mathcal{P}\exp\left[-igt_\alpha^{[i]} v_\mu\int_0^\infty
    ds\, A^\mu_\alpha(\mathbf{b}_T+sv)\right]\,,
  \label{eq:WilsonLine}
\end{equation}
where $t_\alpha^{[i]}$ are the generators of SU(3) in the fundamental
($i=q$) or adjoint ($i=g$) representations. It is interesting to
observe that the soft function in
Eq.~(\ref{eq:softfunctiondefLorentz}) reduces to one at
$\mathbf{b}_T=\mathbf{0}$. This can immediately be seen by plugging
the Wilson line into Eq.~(\ref{eq:WilsonLine}) and using twice the
identity
$W_{v,i}^\dag (\mathbf{0}) W_{v,i} (\mathbf{0})=\mathbb{I}_{N_i\times
  N_i}$ whose trace cancels against the factor $1/N_i$ in
Eq.~(\ref{eq:softfunctiondefLorentz}). We will explicitly verify this
property up to one loop below. This explains why in
$\mathbf{k}_T$-integrated correlators the soft function plays no role.

\begin{figure}[t]
  \centering
  \includegraphics[width=0.95\textwidth]{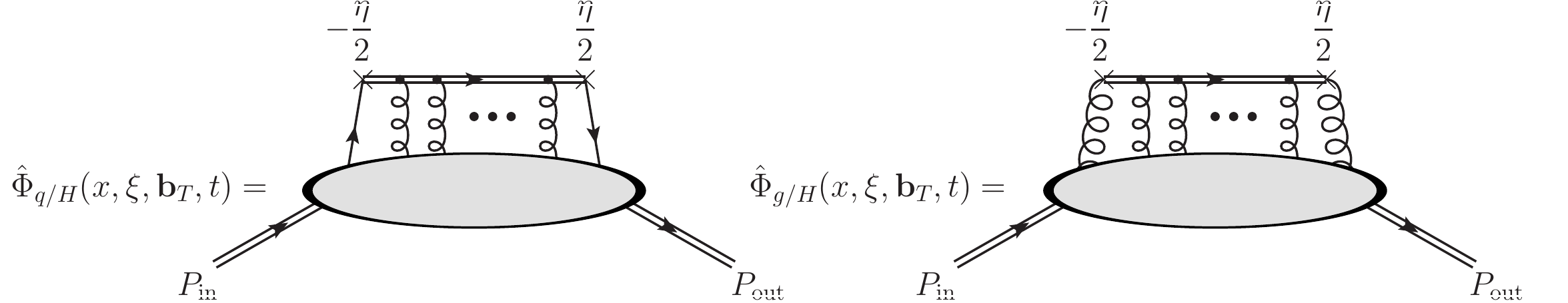}
  \vspace{20pt}
  \caption{Graphical representation of the quark (left) and gluon
    (right) unsubtracted GTMD correlators defined in
    Eq.~(\ref{eq:GTMDdefinition}).\label{fig:HadronGTMDs}}
\end{figure}

A graphical representation of the unsubtracted GTMD correlators
defined in Eq.~(\ref{eq:GTMDdefinition}) is given in
Fig.~\ref{fig:HadronGTMDs}. Here the double lines with an arbitrary
number of gluons attaching to the hadronic blobs represent the
(expansion of the) Wilson line. Notice that the actual Wilson line
that connects the space-time points $-\eta/2$ and $\eta/2$ runs back
and forth along the light-cone direction defined by $n$ (future- or
past-pointing, according to the process) and the two branches are
connected along the transverse direction defined by $\mathbf{b}_T$ at
light-cone infinity. This latter connection is suppressed in Lorentz
gauge and can thus be neglected.

\begin{figure}[t]
  \centering
  \includegraphics[width=0.45\textwidth]{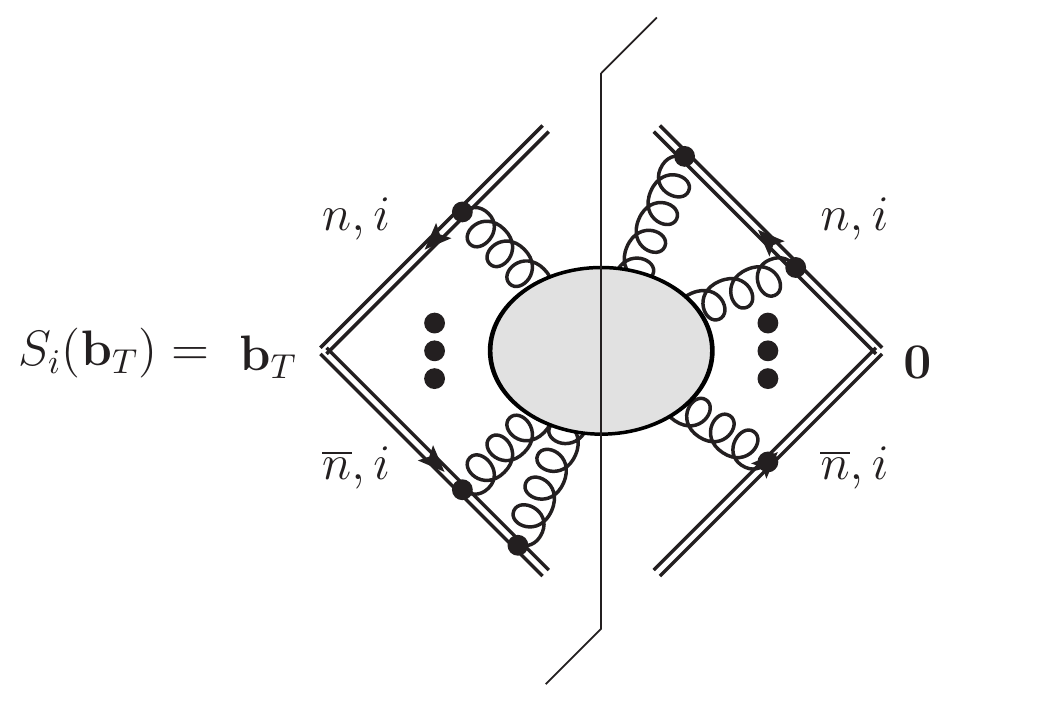}
  \vspace{20pt}
  \caption{Graphical representation of the soft function defined in
    Eq.~(\ref{eq:softfunctiondefLorentz}).\label{fig:SoftFunction}}
\end{figure}

A graphical representation of the soft function defined in
Eq.~(\ref{eq:softfunctiondefLorentz}) is instead given in
Fig.~\ref{fig:SoftFunction}. Two pairs of longitudinal Wilson lines in
$n$ and $\overline{n}$ directions are joint at the origin and at the
space-time point with transverse displacement $\mathbf{b}_T$. An
arbitrary number of gluons is exchanged including loop corrections
represented by the grey blob. Each gluon attachment to the Wilson
lines comes with an SU(3) generator $t_\alpha^{[i]}$ in the
appropriate representation as indicated by the index $i$ associated to
the Wilson lines.

\subsection{Renormalisation and evolution of
  GTMDs}\label{sec:renormalisation_and_evolution}

The subtracted GTMD correlator $\hat{\mathcal{F}}_{i/H}$ in
Eq.~(\ref{eq:subGTMD}), while being free of rapidity singularities, is
UV divergent. In order to renormalise it, we need to separately
renormalise the soft function $\hat{S}_i$ and the unsubtracted GTMD
correlator $\hat{\Phi}_{i/H}$.

The soft function is renormalised multiplicatively as follows:
\begin{equation}
  {S}_i(\mathbf{b}_T,\mu,\zeta,\delta) =\lim_{\epsilon\rightarrow 0}{Z}_{S,i}^{-1}(\mathbf{b}_T,Q, \zeta,\mu,\delta,\epsilon) \hat{S}_i(\mathbf{b}_T,Q,\delta,\epsilon)\,.
  \label{eq:SoftFunctionRenormalisation}
\end{equation}
For completeness, in this expression we have explicitly reported all
the possible dependences that emerge from a perturbative
calculation. Specifically, we have: the scale $Q$ and the regulator
$\delta$ introduced by the regularisation of the rapidity divergences
(to be discussed in Sect.~\ref{sec:SoftFunction}), the scale $\mu$ and
the dimensional regulator $\epsilon$ introduced by the regularisation
of the UV divergences.  Also notice the presence amongst the
dependences of the renormalisation constant ${Z}_{S,i}$ of the
(squared) scale $\zeta$, usually dubbed rapidity scale, also discussed
in Sect.~\ref{sec:SoftFunction}.

The renormalisation of the unsubtracted GTMD correlators is also
purely multiplicative, \textit{i.e.} it does not imply any convolution
nor mixing between quark and gluon operators:
\begin{equation}
  \Phi_{i/H}(x,\xi,\mathbf{b}_T,t,\mu,\delta) =\lim_{\epsilon\rightarrow 0}
  Z_{\Phi,i}^{-1}(\xi,\mu,\delta,\epsilon)\hat{\Phi}_{i/H}(x,\xi,\mathbf{b}_T,t,\delta,\epsilon)\,.
  \label{eq:unsubtractedGTMDRenormalisation}
\end{equation}
This simple pattern is a direct consequence of the fact that the
$\mathbf{b}_T$ displacement prevents any contact divergence of the
operators. Using Eqs.~(\ref{eq:SoftFunctionRenormalisation})
and~(\ref{eq:unsubtractedGTMDRenormalisation}) in
Eq.~(\ref{eq:subGTMD}), one can easily deduce the renormalisation of
the subtracted GTMD correlator:
\begin{equation}
\begin{array}{rcl}
  \displaystyle
  \mathcal{F}_{i/H}(x,\xi,\mathbf{b}_T,t,\mu,\zeta)&=&\displaystyle \lim_{\epsilon,\delta\rightarrow 0}
  {Z}_{S,i}^{1/2}(\mathbf{b}_T,Q, \zeta,\mu,\delta,\epsilon)
  Z_{\Phi,i}^{-1}(\xi,\mu,\delta,\epsilon)
  \hat{\mathcal{F}}_{i/H}(x,\xi,\mathbf{b}_T,t,\delta,\epsilon)\\
\\
&=&\displaystyle \lim_{\delta\rightarrow 0}{S}_i^{-1/2}(\mathbf{b}_T,\mu,\zeta,\delta) \Phi_{i/H}(x,\xi,\mathbf{b}_T,t,\mu,\delta)\,.
\end{array}
\label{eq:renormalisedGTMD}
\end{equation}

Exploiting the fact that, in the limit $\epsilon,\delta\rightarrow 0$,
the bare subtracted GTMD correlator does not depend on either the
renormalisation scale $\mu$ or on the rapidity scale $\zeta$, it is
possible to derive the following evolution equations:
\begin{equation}
  \begin{array}{l}
    \displaystyle
    \frac{d\ln\mathcal{F}_{i/H}(x,\xi,\mathbf{b}_T,t,\mu,\zeta)}{d\ln\sqrt{\zeta}}=
                                                                                      K_i(\mathbf{b}_T,\mu)\,,\\
    \\
    \displaystyle\frac{d\ln\mathcal{F}_{i/H}(x,\xi,\mathbf{b}_T,t,\mu,\zeta)}{d\ln\mu}= \gamma_i(\mu,\zeta)\,.
  \end{array}
  \label{eq:evolutionequations}
\end{equation}
The anomalous dimensions $K_i$ and $\gamma_i$ derive from the
renormalisation constants as follows:
\begin{equation}
  \begin{array}{rcl}
    K_i(\mathbf{b}_T,\mu) &=& \displaystyle \lim_{\epsilon,\delta\rightarrow 0}
  \frac{d\ln{Z}_{S,i}(\mathbf{b}_T,Q,
                              \zeta,\mu,\delta,\epsilon)}{d\ln\zeta}\,,\\
    \\
    \gamma_i(\mu,\zeta) &=&\displaystyle \lim_{\epsilon,\delta\rightarrow 0}\frac{d\ln[{Z}_{S,i}^{1/2}(\mathbf{b}_T,Q, \zeta,\mu,\delta,\epsilon) Z_{\Phi,i}^{-1}(\xi,\mu,\delta,\epsilon)]}{d\ln\mu}\,.
  \end{array}
  \label{eq:evolutionkernel}
\end{equation}
The first equation in Eq.~(\ref{eq:evolutionequations}) is usually
referred to as Collins-Soper equation~\cite{Collins:1981uk,
  Collins:1981va}, while the second is a more common
renormalisation-group equation.

The anomalous dimensions in Eq.~(\ref{eq:evolutionkernel}) can be
mutually related by observing that the cross derivative of the GTMD
correlator must coincide. Indeed, it turns out that:
\begin{equation}
  \frac{d K_i(\mathbf{b}_T,\mu)}{d\ln\mu} = \frac{d
    \gamma_i(\mu,\zeta)}{d\ln\sqrt{\zeta}}\equiv -
  \gamma_{K,i}(a_s(\mu))\,,
  \label{eq:TMDCrossDerivative}
\end{equation}
where the anomalous dimension $\gamma_{K,i}$, usually called cusp
anomalous dimension, only depends on the strong coupling
$a_s=g^2/16\pi^2=\alpha_s/4\pi$. Therefore, it is a purely
perturbative quantity that, for sufficiently large scales, admits the
expansion:
\begin{equation}
  \gamma_{K,i} (a_s(\mu)) = \sum_{n=0}^\infty a_s^{n+1}(\mu)\gamma_{K,i}^{[n]}\,.
\end{equation}

Eq.~(\ref{eq:TMDCrossDerivative}) can be solved w.r.t. both $K_i$ and
$\gamma_i$ expressing these anomalous dimensions in terms of more
fundamental quantities computable in perturbation theory. To do so, we
need appropriate boundary conditions. For $K_i$ the boundary condition
is conveniently set at the scale $\mu=b_0/|\mathbf{b}_T|\equiv \mu_b$,
with $b_0\equiv 2e^{-\gamma_{\rm E}}$ and $\gamma_{\rm E}$ the Euler
constant, where it admits the perturbation expansion:
\begin{equation}
  K_i(\mathbf{b}_T,\mu_b) = \sum_{n=0}^\infty a_s^{n+1}(\mu_b)K_i^{[n]}\,.
\end{equation}
With this boundary condition, the solution to
Eq.~(\ref{eq:TMDCrossDerivative}) reads:
\begin{equation}
  K_i(\mathbf{b}_T,\mu) = K_i(\mathbf{b}_T,\mu_b) - \int_{\mu_b}^{\mu}
  \frac{d\mu'}{\mu'} \gamma_{K,i}(a_s(\mu'))\,.
  \label{eq:CSresummed}
\end{equation}
We notice that, thanks to the integral in the r.h.s., this expression
resums all powers in $a_s$ through its evolution, preventing the
presence of potentially large logarithms. However, as we will
explicitly see below, a perturbative calculation of $K_i$ at a generic
fixed scale $\mu$ produces terms accompanied by powers of
$\ln(\mu/\mu_b)$. Since $\mathbf{b}_T$ is usually integrated over,
these logarithms can become arbitrarily large invalidating any
fixed-order calculation. Therefore, in phenomenological applications
the resummed expression of $K_i$ in Eq.~(\ref{eq:CSresummed}) is to be
preferred over a fixed-order calculation.

We now solve Eq.~(\ref{eq:TMDCrossDerivative}) for $\gamma_i$. In this
case, the boundary condition is conveniently set at
$\sqrt{\zeta} = \mu/\sqrt{1-\xi^2}$ where the anomalous dimension can
be expanded as:
\begin{equation}
  \gamma_i(\mu,\mu/\sqrt{1-\xi^2}) \equiv \gamma_{F,i}(a_s(\mu)) = \sum_{n=0}^\infty a_s^{n+1}(\mu_b)\gamma_{F,i}^{[n]}\,.
\end{equation}
Owing to the fact that the r.h.s. of Eq.~(\ref{eq:TMDCrossDerivative})
does not depend of $\zeta$, the solution to the evolution equation is
particularly simple and reads:
\begin{equation}
  \gamma_i(\mu,\zeta) = \gamma_{F,i}(a_s(\mu)) - \gamma_{K,i}(a_s(\mu))\ln\left(\frac{\sqrt{(1-\xi^2)\zeta}}{\mu}\right)\,.
\end{equation}

In conclusion, the evolution equations for the GTMD correlator take
the explicit form:
\begin{equation}
  \begin{array}{rcl}
    \displaystyle
    \frac{d\ln\mathcal{F}_{i/H}(x,\xi,\mathbf{b}_T,t,\mu,\zeta)}{d\ln\sqrt{\zeta}}&=&\displaystyle
                                                                                      K_i(\mathbf{b}_T,\mu_b) - \int_{\mu_b}^{\mu}
                                                                                      \frac{d\mu'}{\mu'} \gamma_{K,i}(a_s(\mu'))\,,\\
    \\
    \displaystyle\frac{d\ln\mathcal{F}_{i/H}(x,\xi,\mathbf{b}_T,t,\mu,\zeta)}{d\ln\mu}&=&\displaystyle \gamma_{F,i}(a_s(\mu)) - \gamma_{K,i}(a_s(\mu))\ln\left(\frac{\sqrt{(1-\xi^2)\zeta}}{\mu}\right)\,,
  \end{array}
  \label{eq:evolutionequationsExp}
\end{equation}
where the relevant anomalous dimensions $K_i(\mathbf{b}_T,\mu_b)$,
$\gamma_{F,i}$, and $\gamma_{K,i}$ are all perturbative quantities. As
a by-product of the calculation of the matching functions presented
below, we will extract the leading-term coefficients $K_i^{[0]}$,
$\gamma_{F,i}^{[0]}$, and $\gamma_{K,i}^{[0]}$. Unsurprisingly, they
turn out to be identical to those obtained in TMD factorisation.

\subsection{Matching on GPDs}\label{sec:matching}

As mentioned above, the transverse displacement $\mathbf{b}_T$ is
Fourier conjugated to the partonic transverse momentum
$\mathbf{k}_T$. As a consequence, when
$\mathbf{b}_T\simeq \mathbf{0}$, which corresponds to the emission of
partons with large $\mathbf{k}_T$, it is legitimate to expect such
emissions to be treatable in perturbation theory. As a matter of fact,
in this regime hard emissions can be factorised into perturbatively
calculable quantities, the matching functions, that allow one to
express the GTMD correlators in terms of the corresponding collinear
GPD correlators. The matching formula reads:
\begin{equation}
  \begin{array}{rcl}
    \displaystyle \mathcal{F}_{i/H}(x,\xi,\mathbf{b}_T,t,\mu,\zeta) &=&\displaystyle 
                                                                        \int_x^\infty\frac{dy}{y}
                                                                        \mathcal{C}_{i/k}\left(y,\frac{\xi}{x},\mathbf{b}_T,\mu,\zeta\right)F_{k/H}\left(\frac{x}{y},\xi,t,\mu\right)\\
    \\
                                                                    &\equiv&\displaystyle
                                                                             \mathcal{C}_{i/k}(x,\kappa,\mathbf{b}_T,\mu,\zeta)\mathop{\otimes}_{x} F_{k/H}\left(x,\xi,t,\mu\right)\,,
    \end{array}
\label{eq:matchingOnToGPDs}
\end{equation}
where $F_{k/H}$, with $k=q,g$, are the renormalised GPD correlators
whose explicit definition can be found, \textit{e.g.}, in
Ref.~\cite{Bertone:2022frx}. In addition, we have defined
$\kappa\equiv\xi/x$.

In order to extract the matching functions $\mathcal{C}_{i/k}$, we
make use of the concept of parton-in-parton
distribution~\cite{Collins:2011zzd} (sometimes also referred to as
quark-target model, see \textit{e.g.} Ref.~\cite{Goeke:2006ef}). The
idea is to replace the hadronic states involved in the correlators in
Eq.~(\ref{eq:GTMDdefinition}) and in their collinear analogues with
partonic states. Since the action of partonic fields on partonic
states is computable in perturbation theory, this enables a direct
calculation. It should be pointed out that the validity of this
procedure is tightly connected to factorisation and the universality
of the resulting partonic distributions~\cite{Collins:2011zzd}. Since
so far, to the best of my knowledge, no actual factorisation theorems
involving GTMDs have been \textit{proven},\footnote{However,
  factorisation is often assumed. See, for example,
  Refs.~\cite{Hatta:2016dxp, Bhattacharya:2017bvs,
    Bhattacharya:2018lgm, Boer:2021upt}.} here we simply assume its
applicability. The parton-in-parton version of
Eq.~(\ref{eq:matchingOnToGPDs}) is then:
\begin{equation}
  \mathcal{F}_{i/j}(x,\xi,\mathbf{b}_T,\mu,\zeta) =
  \mathcal{C}_{i/k}(x,\kappa,\mathbf{b}_T,\mu,\zeta)\mathop{\otimes}_{x} F_{k/j}\left(x,\xi,\mu\right)\,,
\label{eq:matchingOnToGPDsParton}
\end{equation}
where $j=q,g$. Notice that we have dropped the dependence on $t$ that
does not participate in a partonic computation. Now all quantities
involved in the matching formula admit a perturbative expansion:
\begin{equation}
  \begin{array}{rcl}
    \displaystyle \mathcal{F}_{i/j}(x,\xi,\mathbf{b}_T,\mu,\zeta)
    &=&\displaystyle
        \sum_{n=0}^{\infty}a_s^n\mathcal{F}_{i/j}^{[n]}(x,\xi,\mathbf{b}_T,\mu,\zeta)\,,\\
    \\
    \displaystyle {F}_{k/j}(x,\xi,\mu)
    &=&\displaystyle
        \sum_{n=0}^{\infty}a_s^n{F}_{k/j}^{[n]}(x,\xi,\mu) \,,\\
    \\
    \displaystyle \mathcal{C}_{i/k}(x,\kappa,\mathbf{b}_T,\mu,\zeta)
    &=&\displaystyle
        \sum_{n=0}^{\infty}a_s^n\mathcal{C}_{i/k}^{[n]}(x,\kappa,\mathbf{b}_T,\mu,\zeta) \,.
  \end{array}
\end{equation}

At the lowest order in $a_s$, where no additional radiation is
allowed, one finds that:
\begin{equation}
  \mathcal{F}_{i/j}^{[0]}(x,\xi,\mathbf{b}_T,\mu,\zeta)={F}_{i/j}^{[0]}(x,\xi,\mu)=D_j(\xi)\delta_{ij}\delta(1-x)\,,
  \label{eq:LOunsubtractedGTMDs}
\end{equation}
with $D_q(\xi)=\sqrt{1-\xi^2}$ and
$D_g(\xi)=1-\xi^2$~\cite{Bertone:2022frx}. This immediately implies
that:
\begin{equation}\label{eq:LOmatching}
  \mathcal{C}_{i/k}^{[0]}(x,\kappa,\mathbf{b}_T,\mu,\zeta)=\delta_{ik}\delta(1-x)\,.
\end{equation}

These results allow us to determine the one-loop ($\mathcal{O}(a_s)$)
correction to $ \mathcal{C}_{i/j}$ in terms of the one-loop GTMD and
GPD parton-in-parton correlators:
\begin{equation}
 \mathcal{C}_{i/j}^{[1]}(x,\kappa,\mathbf{b}_T,\mu,\zeta)= D_j^{-1}(\xi)\left[\mathcal{F}_{i/j}^{[1]}(x,\xi,\mathbf{b}_T,\mu,\zeta)-F_{i/j}^{[1]}\left(x,\xi,\mu\right)\right]\,.
\label{eq:matchingOnToGPDsParton}
\end{equation}
Using the perturbative expansion of the renormalised soft function
$S_i$ and unsubtracted parton-in-parton GTMD correlator
${\Phi}_{i/j}$:
\begin{equation}
\begin{array}{rcl}
  \displaystyle
  {S}_i(\mathbf{b}_T,\mu,\zeta,\delta) &=&\displaystyle
                                           \sum_{n=0}^{\infty} a_s^n {S}_i^{[n]}(\mathbf{b}_T,\mu,\zeta,\delta)\,,\\
  \\
  \Phi_{i/j}(x,\xi,\mathbf{b}_T,\mu,\delta) &=&\displaystyle \sum_{n=0}^{\infty} a_s^n \Phi_{i/j}^{[n]}(x,\xi,\mathbf{b}_T,\mu,\delta)\,,
\end{array}
\label{eq:RenSandPhiExpansions}
\end{equation}
with ${S}_i^{[0]}(\mathbf{b}_T,\mu,\zeta,\delta)=1$ (see
Eq.~(\ref{eq:LOSF})) in Eq.~(\ref{eq:renormalisedGTMD}), we obtain:
\begin{equation}
\mathcal{F}_{i/j}^{[1]}(x,\xi,\mathbf{b}_T,\mu,\zeta) =
\Phi_{i/j}^{[1]}(x,\xi,\mathbf{b}_T,\mu,\delta) - \frac12 D_j(\xi)\delta_{ij}\delta(1-x) {S}_i^{[1]}(\mathbf{b}_T,\mu,\zeta,\delta)\,.
\end{equation}
Notice that, while the single terms on r.h.s. of this equation depend
on the rapidity regulator $\delta$, the l.h.s. does not. This means
that this dependence must cancel in this specific combination
confirming at one loop the rapidity-divergence safety of the
definition in Eq.~(\ref{eq:subGTMD}). Plugging this identity into
Eq.~(\ref{eq:matchingOnToGPDsParton}), finally gives:
\begin{equation}
\begin{array}{rcl}
  \displaystyle \mathcal{C}_{i/k}^{[1]}(x,\kappa,\mathbf{b}_T,\mu,\zeta)&=&\displaystyle 
                                                                            D_k^{-1}(\xi)\left[\Phi_{i/k}^{[1]}(x,\xi,\mathbf{b}_T,\mu,\delta)
                                                                            -F_{i/k}^{[1]}\left(x,\xi,\mu\right) \right] \\
  \\
                                                                        &-&\displaystyle  \frac12
                                                                            \delta_{ik}\delta(1-x){S}_i^{[1]}(\mathbf{b}_T,\mu,\zeta,\delta)\,.
\end{array}
\label{eq:matchingOnToGPDsPartonExpanded}
\end{equation}
Therefore, the computation of the one-loop corrections to the matching
functions boils down to computing parton-in-parton unsubtracted GTMD
and GPD correlators, and the soft function at the same
order. Moreover, the computations of $\Phi_{i/k}^{[1]}$ and of
$F_{i/k}^{[1]}$ are closely related, with the latter recently
presented in Ref.~\cite{Bertone:2022frx}. This similarity can be
exploited to simplify the calculation. First of all, we notice that
they are both made of a ``real'' and a ``virtual''
contribution:\footnote{Strictly speaking, since both GTMDs and GPDs
  are involved in exclusive processes, all contributions are
  virtual. However, one can distinguish between contributions that
  only affect either the left or the right part of the operator
  (\textit{i.e.} that live at at $-\eta/2$ \textit{or} $\eta/2$, with
  reference to Fig.~\ref{fig:HadronGTMDs}), defined as virtual, and
  contributions that instead connect the two parts, defined as
  real. This is done in analogy with the inclusive case in which the
  final-state cut sets the real contributions on shell.}
\begin{equation}
\Phi_{i/k}^{[1]}=\Phi_{i/k}^{[1],\rm real}+\Phi_{i/k}^{[1],\rm
  virt}\,,\qquad  F_{i/k}^{[1]} = F_{i/k}^{[1],\rm real} +
F_{i/k}^{[1],\rm virt}\,.
\end{equation}
Since the virtual contribution to the GTMD correlator in
$\mathbf{k}_T$ space must, for kinematic reasons, be proportional to
$\delta^{(2)}(\mathbf{k}_T)$, its Fourier transform is independent
from $\mathbf{b}_T$. But this is precisely the same kinematics used to
compute the virtual contribution to the GPD correlator. Therefore, one
finds:
\begin{equation}
\Phi_{i/k}^{[1],\rm virt} = F_{i/k}^{[1],\rm virt}\,,
\label{eq:vitualequals}
\end{equation}
such that:
\begin{equation}
  \Phi_{i/k}^{[1]}-F_{i/k}^{[1]} = \Phi_{i/k}^{[1],\rm real}
  -F_{i/k}^{[1], \rm real}\,.
\label{eq:differencereals}
\end{equation}
To compute the difference in the r.h.s., we can again use the results
of Ref.~\cite{Bertone:2022frx}. To do so, we observe that the only
difference between $F_{i/k}^{[1], \rm real}$ and
$\Phi_{i/k}^{[1],\rm real}$ in $\mathbf{b}_T$ space is that the
$\mathbf{k}_T$ integral in the latter is weighted by the phase
$e^{i \mathbf{b}_T\cdot \mathbf{k}_T}$, reflecting the displacement of
the partonic fields. Specifically, one finds:
\begin{equation}
  \Phi_{i/k}^{[1],\rm real}(x,\xi,\mathbf{b}_T,\mu,\delta) =
  D_k(\xi)\left[\mathcal{P}_{i/k}^{[0], \rm real}(x,\kappa,\delta)-\epsilon
    \mathcal{R}_{i/k}^{[1]}(x,\kappa)\right] \mu^{2\epsilon} 4\pi
  \int\frac{d^{2-2\epsilon}\mathbf{k}_T}{(2\pi)^{2-2\epsilon}}\frac{e^{i\mathbf{b}_T\cdot
      \mathbf{k}_T}}{\mathbf{k}_T^2} \,,
\label{eq:GTMDreal}
\end{equation}
where $\mathcal{P}_{i/k}^{[0], \rm real}$ is the real (and rapidity
divergent) contribution to the one-loop GPD splitting functions, while
the ``residual'' functions $\mathcal{R}_{i/k}^{[1]}$ were so far
unknown and are computed here for the first time (see
Sect.~\ref{sec:unsubtractedGTMDs}). Importantly, the exponential in
the integral in Eq.~(\ref{eq:GTMDreal}) regulates the UV divergence
for $\mathbf{k}_T\rightarrow \infty$~\cite{Bacchetta:2008xw}. As a
consequence, the UV divergences of the GTMD correlator can only be in
the virtual (diagonal) part, hence the simple renormalisation pattern
in Eq.~(\ref{eq:unsubtractedGTMDRenormalisation}).

The $\mathbf{k}_T$ integral in Eq.~(\ref{eq:GTMDreal}) can be computed
analytically and evaluates to:
\begin{equation}
  4\pi
  \int\frac{d^{2-2\epsilon}\mathbf{k}_T}{(2\pi)^{2-2\epsilon}}\frac{e^{i\mathbf{b}_T\cdot
      \mathbf{k}_T}}{\mathbf{k}_T^2}
  =\pi^\epsilon b_T^{2\epsilon}\Gamma(-\epsilon)=
  -\frac{\pi^\epsilon b_T^{2\epsilon}(1+\gamma_{\rm
      E}\epsilon)}{\epsilon_{\rm IR}}+\mathcal{O}(\epsilon)\,,
\end{equation}
where $b_T\equiv|\mathbf{b}_T|$.  In the rightmost equality we have
highlighted the fact that the pole for $\epsilon\rightarrow 0$ is of
infrared origin. This divergence is cancelled in the difference in
Eq.~(\ref{eq:differencereals}) by an analogous divergence in the
UV-renormalised GPD correlator. To see this, we observe that the real
part of the renormalised GPD correlator can be written as:
\begin{equation}
  F_{i/k}^{[1],\rm real}(x,\xi,\mu) =D_k(\xi)
  S_\epsilon\left[\mathcal{P}_{i/k}^{[0], \rm
      real}(x,\kappa,\delta)\left(\ln\mu^2-\frac{\mu^{2\epsilon}}{\epsilon_{\rm
          IR}}\right)-\epsilon\mathcal{R}_{i/k}^{[1]}(x,\kappa)\left(\frac{\mu^{2\epsilon}}{\epsilon_{\rm
          UV}}-\frac{\mu^{2\epsilon}}{\epsilon_{\rm IR}}\right)\right]
  \,,
\label{eq:collGPDren}
\end{equation}
where:
\begin{equation}
  S_\epsilon=\frac{(4\pi)^\epsilon}{\Gamma(1-\epsilon)} = 1 +
  \epsilon\left(\ln 4\pi-\gamma_{\rm
      E}\right)+\mathcal{O}(\epsilon^2)\,.
\label{eq:Sepsilon}
\end{equation}
From Eq.~(\ref{eq:collGPDren}), it is apparent that the UV divergence
has been removed leaving a $\ln\mu^2$, while the infrared divergence
is still present. In addition, we also retained the next term in the
expansion in powers of $\epsilon$. This term, proportional to
$\mathcal{R}_{i/k}^{[1]}$, multiplies a scaleless integral in which UV
($1/\epsilon_{\rm UV}$) and IR ($1/\epsilon_{\rm IR}$) poles cancel
giving a vanishing result~\cite{Collins:1984xc} that does not
contribute to $F_{i/k}^{[1],\rm real}$. However, since the IR
contribution is cancelled by the GPD correlator, the UV pole does give
a finite contribution in the combination in
Eq.~(\ref{eq:differencereals}). Indeed, using Eqs.~(\ref{eq:GTMDreal})
and~(\ref{eq:collGPDren}), Eq.~(\ref{eq:differencereals}) yields:
\begin{equation}
\Phi_{i/k}^{[1]}(x,\xi,\mathbf{b}_T,\mu,\delta)  -F_{i/k}^{[1]}(x,\xi,\mu) = D_k(\xi)\left[-\mathcal{P}_{i/k}^{[0],\rm
    real}(x,\kappa,\delta) \ln\left(\frac{\mu^2}{\mu_b^2}\right)+\mathcal{R}_{i/k}^{[1]}(x,\kappa)\right]\,,
\label{eq:differencerealsexplicit}
\end{equation}
where the limit $\epsilon\rightarrow 0$ has already been taken using
the equality:
\begin{equation}
  \lim_{\epsilon\rightarrow
    0}\frac{S_\epsilon-\pi^\epsilon b_T^{2\epsilon}(1+\gamma_{\rm E}\epsilon)}{\epsilon}
  =\ln\frac{4e^{-2\gamma_{\rm E}}}{b_T^{2}}=\ln\frac{b_0^2}{b_T^2}=\ln\mu_b^2\,.
\end{equation}
Remarkably, all divergences have cancelled leaving a finite
quantity. Plugging Eq.~(\ref{eq:differencerealsexplicit}) into
Eq.~(\ref{eq:matchingOnToGPDsPartonExpanded}), we obtain:
\begin{equation}
  \mathcal{C}_{i/k}^{[1]}(x,\kappa,\mathbf{b}_T,\mu,\zeta)= -\mathcal{P}_{i/k}^{[0],\rm
    real}(x,\kappa,\delta) \ln \left(\frac{\mu^2}{\mu_b^2}\right)+\mathcal{R}_{i/k}^{[1]}(x,\kappa) - \frac12 \delta_{ik}\delta(1-x){S}_i^{[1]}(\mathbf{b}_T,\mu,\zeta,\delta)\,.
\label{eq:semifinalres}
\end{equation}

It is now convenient to extract the rapidity divergence from
$\mathcal{P}_{i/k}^{[0],\rm real}$ in a way to facilitate the
cancellation against the soft function ${S}_i^{[1]}$. To this purpose,
we now introduce the particular strategy that will be used to regulate
the rapidity divergences. We will employ the principal-value
prescription~\cite{Curci:1980uw} that acts on eikonal propagators of
the kind $1/(n\cdot k)$ and $1/(\overline{n}\cdot k)$ by replacing
them with their principal-valued version, that is:
\begin{equation}
  \frac{1}{(n\cdot k)}\rightarrow {\rm PV}\frac{1}{(n\cdot k)}=\frac{1}{2}\left[\frac{1}{(n\cdot k)+i\delta (n\cdot p)}+\frac{1}{(n\cdot k)-i\delta
      (n\cdot p)}\right]=\frac{(n\cdot k)}{(n\cdot k)^2+\delta^2 (n\cdot p)^2}\,,
\label{eq:PVprescription}
\end{equation}
with $\delta\rightarrow 0$, and analogously for
$1/(\overline{n}\cdot k)$ with $p$ replaced by $\overline{p}$. The
momenta $p$ and $\overline{p}$ are to be thought as the light-like
momenta ($p^2=\overline{p}^2=0$) of two highly energetic projectiles
that collide with large centre-of-mass energy
$(p+\overline{p})^2=2p\cdot\overline{p}\equiv Q^2\gg \Lambda_{\rm
  QCD}^2$. $Q$ is precisely the scale first introduced in
Eq.~(\ref{eq:SoftFunctionRenormalisation}) as a consequence of the
regularisation of rapidity divergences and that will appear explicitly
only in the soft function. When integrating over the partonic momentum
$k$, the longitudinal projection $(n\cdot k)$ is usually parameterised
as $(n\cdot k)=(1-z) (n\cdot p)$, with the variable $z$ running in the
interval $z\in[0,1]$. One can then show that the principal-value
prescription is equivalent to replacing:
\begin{equation}
\frac{1}{1-z}\rightarrow
\left(\frac{1}{1-z}\right)_+-\delta(1-z) \ln\delta\,,
\label{eq:deltaregularisation}
\end{equation}
where the $+$-prescription is defined as:
\begin{equation}
\int_0^1dz\left(\frac{1}{1-z}\right)_+f(z) = \int_0^1\frac{f(z)-f(1)}{1-z}\,,
\end{equation}
for a test function $f$ well-behaved at $z=1$.

With this at hand, we can replace $\mathcal{P}_{i/k}^{[0],\rm real}$
in Eq.~(\ref{eq:semifinalres}) with the full splitting function
$\mathcal{P}_{i/k}^{[0]}$ using the following equality:
\begin{equation}
\begin{array}{rcl}
  \mathcal{P}_{i/k}^{[0],\rm real}(x, \kappa,\delta) &=&\displaystyle \mathcal{P}_{i/k}^{[0]}(x, \kappa)
  -\mathcal{P}_{i/k}^{[0],\rm virt}(x, \kappa,\delta)\\
\\
&=&\displaystyle \mathcal{P}_{i/k}^{[0]}(x,\kappa)
  -\delta_{ik}\delta(1-x)2C_i\left[K_i-\ln(1-\xi^2)-2\int_0^1\frac{dz}{1-z}\right]\\
\\
&=&\displaystyle \mathcal{P}_{i/k}^{[0]}(x,\kappa)
  -\delta_{ik}\delta(1-x)2C_i\left[K_i-\ln(1-\xi^2)+2\ln\delta\right]\,.
  \end{array}
\label{eq:replacement}
\end{equation}
In the second line we have used the explicit form of the virtual
contribution to the splitting function extracted from
Ref.~\cite{Bertone:2022frx}. The color factors $C_i$ are defined as:
\begin{equation}
  C_g=C_A=N_c=3\,,\qquad C_q=C_F=\frac{N_c^2-1}{2N_c}=\frac{4}{3}\,,
\end{equation}
and the values of the coefficient $K_i$ are:
\begin{equation}
K_q = \frac32\,,\qquad K_g = \frac{11C_A-4n_fT_R}{6C_A}\,,
\end{equation}
with $T_R=1/2$ and $n_f$ the number of active quark flavours. Finally,
in the third line of Eq.~(\ref{eq:replacement}) we have used
Eq.~(\ref{eq:deltaregularisation}) to regularise the divergent
integral in $z$. We can now plug Eq.~(\ref{eq:replacement}) into
Eq.~(\ref{eq:semifinalres}) to obtain:
\begin{equation}
\begin{array}{rcl}
  \mathcal{C}_{i/k}^{[1]}(x,\kappa,\mathbf{b}_T,\mu,\zeta)&=&\displaystyle 
  -\mathcal{P}_{i/k}^{[0]}(x,\kappa)
  \ln\left(\frac{\mu^2}{\mu_b^2}\right)
  +\mathcal{R}_{i/k}^{[1]}(x,\kappa)\\
\\
  &+&\displaystyle\delta_{ik}\delta(1-x)\bigg[2C_i\left(K_i-\ln(1-\xi^2)+2\ln\delta\right)\ln\left(\frac{\mu^2}{\mu_b^2}\right)\\
\\
&-&\displaystyle \frac12  {S}_i^{[1]}(\mathbf{b}_T,\mu,\zeta,\delta)\bigg]\,.
\end{array}
\label{eq:finalres}
\end{equation}
Since the splitting functions $\mathcal{P}_{i/k}^{[0]}$ have been
computed in Ref.~\cite{Bertone:2022frx}, all we are left with to
obtain the one-loop correction to the matching functions is to compute
the soft function ${S}_i^{[1]}$ and the residual functions
$\mathcal{R}_{i/k}^{[1]}$. This will be done next in
Sects.~\ref{sec:SoftFunction} and~\ref{sec:unsubtractedGTMDs},
respectively.


\subsection{The soft function}\label{sec:SoftFunction}

Many calculations of the soft function to one-loop order and beyond
exist in the literature. For example, one-loop results can be found in
Refs.~\cite{Echevarria:2011epo, Vladimirov:2014aja,
  Echevarria:2015byo, Li:2016axz, Ebert:2019okf, Deng:2022gzi}. As of
today, also two-loop~\cite{Li:2011zp, Echevarria:2015byo, Li:2016axz}
and three-loop~\cite{Li:2016ctv, Ebert:2020yqt} results have been
presented. As mentioned above, the soft function is affected by
rapidity divergences. They are caused by the presence of eikonal
propagators like $1/(n\cdot k)$ and cannot be regulated by means of
dimensional regularisation. Therefore, an additional regulator needs
to be introduced. We will use the principal-value
prescription~\cite{Curci:1980uw} introduced in the previous section
that, to the best of my knowledge, is used here for the first time to
compute the soft function. The soft function is also affected by UV
divergences that we regularise by means of dimensional regularisation.

The ultimate goal of this section is to compute the first two terms,
$n=0,1$, of the series in Eq.~(\ref{eq:RenSandPhiExpansions}) for the
renormalised soft function. To do so, we start from the perturbative
series of the \textit{bare} soft function:
\begin{equation}
  \hat{S}_i(\mathbf{b}_T,Q,\delta,\epsilon) =
  \sum_{n=0}^{\infty} a_s^n
  \hat{S}_i^{[n]}(\mathbf{b}_T,Q,\delta,\epsilon)\,.
  \label{eq:SoftFunctionExpansion}
\end{equation}

The leading-order, $n=0$, is trivial. To compute it, we simply
approximate all the Wilson lines in
Eq.~(\ref{eq:softfunctiondefLorentz}) with the unity operator in the
appropriate colour-group representation. This immediately gives:
\begin{equation}
  \hat{S}_i^{[0]}(\mathbf{b}_T,Q,\delta,\epsilon) =
  \frac{1}{N_i}\mbox{Tr}_c\langle0| \mathbb{I}_{N_i\times
    N_i}\mathbb{I}_{N_i\times N_i}\mathbb{I}_{N_i\times
    N_i}\mathbb{I}_{N_i\times N_i}|0\rangle=1\,.
  \label{eq:LOSF}
\end{equation}

\begin{figure}[t]
  \begin{centering}
    \includegraphics[width=0.7\textwidth]{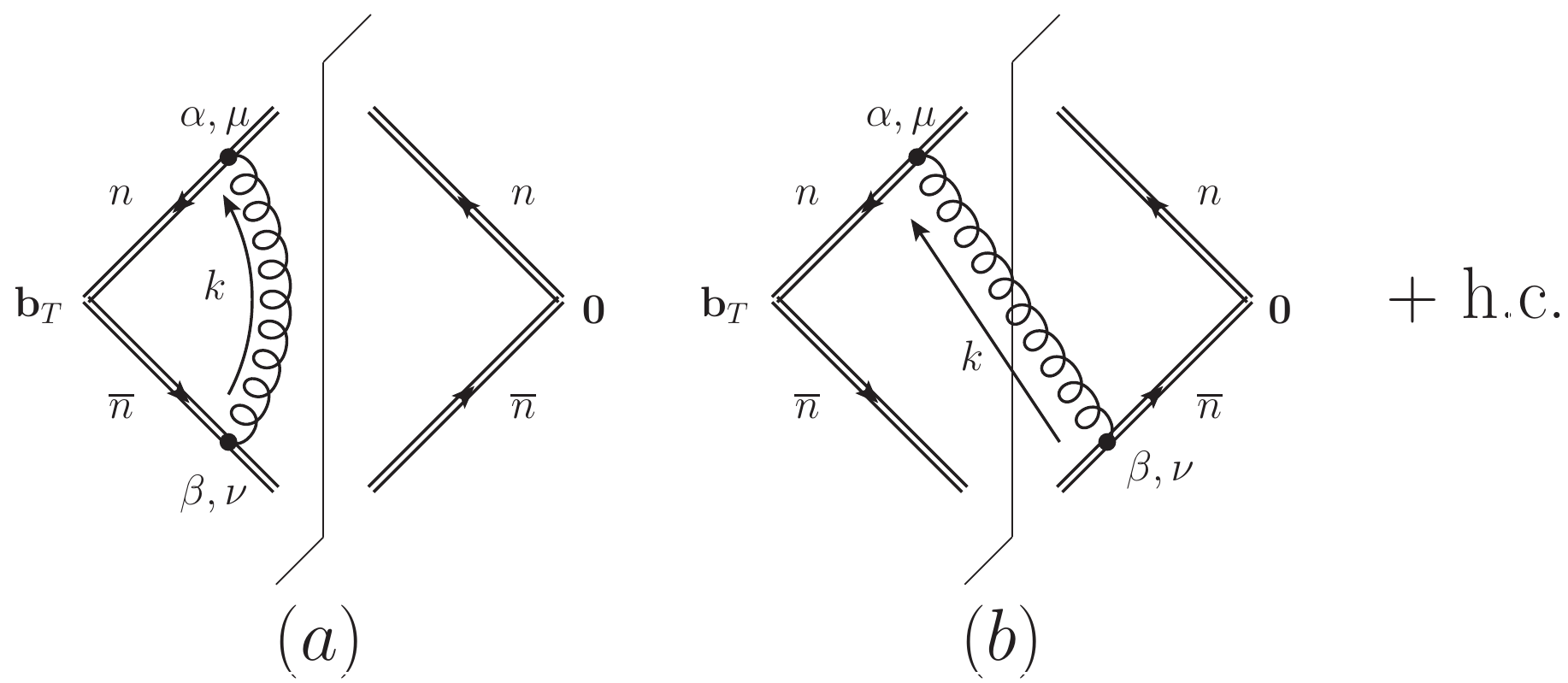}
    \vspace{20pt}
    \caption{Non-vanishing one-loop diagrams contributing to the soft
      function. The vertical bars indicate the final-state cuts that
      set all parton that they cross on the mass
      shell.\label{fig:NLOSF}}
  \end{centering}
\end{figure}

We now move on to computing the one-loop correction, \textit{i.e.} the
term $n=1$ in the series in
Eq.~(\ref{eq:SoftFunctionExpansion}). One-loop diagrams in which the
gluon attaches to Wilson lines pointing in the same light-cone
direction are proportional to either $n^2=0$ or $\overline{n}^2=0$,
and thus give no contribution. On the contrary, the diagrams displayed
in Fig.~\ref{fig:NLOSF} have a gluon that attaches to Wilson lines
pointing in different directions. Therefore, they are proportional to
$n\cdot\overline{n}=1$, giving a non-null contribution.
Applying standard Feynman rules in Lorentz
gauge~\cite{Collins:2011zzd} and using the principal-value
prescription to regulate the eikonal propagators, diagrams (a) and (b)
along with their respective hermitian conjugates separately evaluate
to:
\begin{equation}
  \begin{array}{rcl}
    \hat{S}_i^{[a+a^\dag]} &=& \displaystyle
                               - a_s4 C_i (4\pi\mu^2)^{\epsilon}\frac{1+\cos(\pi\epsilon)}{2}\left(Q^2\delta^2\right)^{-\epsilon}\Gamma^2(\epsilon)
                               \Gamma(1-\epsilon)\,,\\
    \\
    \hat{S}_i^{[b+b^\dag]} &=&\displaystyle - a_s4C_i
                               (4\pi\mu^2)^{\epsilon}
                               \Gamma(-\epsilon) \bigg[\left(\frac{b_T^{2}}{4}\right)^\epsilon\left(\ln\frac{Q^2
                               \delta^2}{\mu_b^2}-\psi(-\epsilon)-\gamma_{\rm
                               E}\right)\\
    \\
                           &-&\displaystyle \frac{1+\cos(\pi\epsilon)}{2}\left(Q^2\delta^2\right)^{-\epsilon}\Gamma(\epsilon)
                               \Gamma(1-\epsilon)\bigg]\,.
  \end{array}
\end{equation}
The net result for the one-loop correction to the bare soft function
is:
\begin{eqnarray}
\nonumber    \hat{S}_i^{[1]}(\mathbf{b}_T,Q,\delta,\epsilon)   &=&a_s^{-1}\left[\hat{S}_i^{[a+a^\dag]}+\hat{S}_i^{[b+b^\dag]}\right]\\
\nonumber    \\
\label{eq:OneLoopSF}                                                      &=&\displaystyle - 4C_i
                                                          (4\pi\mu^2)^{\epsilon}
                                                          \Gamma(-\epsilon) \left(\frac{b_T^{2}}{4}\right)^\epsilon\left(\ln\frac{Q^2
                                                          \delta^2}{\mu_b^2}-\psi(-\epsilon)-\gamma_{\rm
                                                          E}\right)\\
\nonumber    \\
\nonumber                                                      &=&\displaystyle  4C_i\left(-\frac{S_\epsilon^2}{\epsilon^2}+\frac12\ln^2\left(\frac{\mu^2}{\mu_b^2}\right)-\left(\frac{S_\epsilon}{\epsilon}+\ln\left(\frac{\mu^2}{\mu_b^2}\right)\right) \ln\left(\frac {\mu^2}{Q^2\delta^2}\right)+\frac{\pi^2}{12} +\mathcal{O}(\epsilon)\right)\,.
\end{eqnarray}
This result agrees with that of Ref.~\cite{Echevarria:2015byo} despite
the different (but closely related) regularisation of the rapidity
divergences.

A couple of additional comments are in order. First, from the second
line of Eq.~(\ref{eq:OneLoopSF}) we immediately see that
$\hat{S}_i^{[1]}(\mathbf{b}_T =\mathbf{0},Q,\delta,\epsilon)=0$. This
reflects the fact that the soft function must unitarise for
$\mathbf{b}_T =\mathbf{0}$. Considering the leading-order result in
Eq.~(\ref{eq:LOSF}), this requirement is indeed fulfilled to one-loop
accuracy by our calculation. It is also interesting to observe that
$\hat{S}_i^{[b+b^\dag]}$ is UV finite:
\begin{equation}
  \lim_{\epsilon\rightarrow 0}\hat{S}_i^{[b+b^\dag]}(\mathbf{b}_T,Q,\mu,\delta,\epsilon) = 4C_i\left(\frac12\ln^2\left(\frac{Q^2\delta^2 }{\mu_b^2}\right) + \frac{\pi^2}{12}\right)\,.
\end{equation}
This was to be expected because the transverse displacement
$\mathbf{b}_T$ regulates the UV divergence in real diagrams. This is a
further consistency check of the calculation.

We can now renormalise the soft function. This is done
multiplicatively as in
Eq.~(\ref{eq:SoftFunctionRenormalisation}). Considering the first two
orders of the bare soft function, Eqs.~(\ref{eq:LOSF})
and~(\ref{eq:OneLoopSF}), the renormalisation constant ${Z}_{S,i}$ in
the $\overline{\mbox{MS}}$ scheme reads:\footnote{Notice that in the
  $\overline{\mbox{MS}}$ scheme poles in $S_\epsilon/\epsilon$, rather
  than in $1/\epsilon$, are subtracted. The expansion in the second
  line of Eq.~(\ref{eq:OneLoopSF}) is purposely written in terms of
  $S_\epsilon/\epsilon$ in view of renormalisation in the
  $\overline{\mbox{MS}}$ scheme.}
\begin{equation}
{Z}_{S,i}(\mathbf{b}_T,Q, \zeta,\mu,\delta,\epsilon)= 1 - a_s
 4C_i\left[\frac{S_\epsilon^2}{\epsilon^2}+\frac{S_\epsilon}{\epsilon}\ln\left(\frac{\mu^2}{Q^2\delta^2}\right)+\ln\left(\frac{\mu^2}{\mu_b^2}\right)\ln\left(\frac{\zeta}{Q^2}\right)\right]+\mathcal{O}(a_s^2)\,.
 \label{eq:ZSoftFunction}
\end{equation}
Here, the arbitrary rapidity scale $\zeta$ is introduced as a proxy to
parameterise finite $\mathcal{O}(a_s)$ contributions. Using this
renormalisation constant, one obtains the first two coefficients of
the perturbative expansion of the renormalised soft function:
\begin{equation}
\begin{array}{rcl}
  \displaystyle {S}_i^{[0]}(\mathbf{b}_T,\mu,\zeta,\delta) &=& \displaystyle  1\,,\\
  \\
  \displaystyle  {S}_i^{[1]}(\mathbf{b}_T,\mu,\zeta,\delta) &=& \displaystyle 2
                                                                C_i\left(4\ln\left(\frac{\mu^2}{\mu_b^2}\right)\ln\delta+\ln^2\left(\frac{\mu^2}{\mu_b^2}\right)-2\ln\left(\frac{\mu^2}{\mu_b^2}\right)\ln\left(\frac
                                                                {\mu^2}{\zeta}\right)+\frac{\pi^2}{6}\right)\,.
\end{array}
\end{equation}

Evidently, ${S}_i^{[1]}$ is still affected by a rapidity divergence
for $\delta\rightarrow 0$. This divergence is precisely what is needed
to cancel an analogous divergence of the unsubtracted GTMD
correlator. Indeed, plugging ${S}_i^{[1]}$ into
Eq.~(\ref{eq:finalres}), one finds:
\begin{eqnarray}
\nonumber \mathcal{C}_{i/k}^{[1]}(x,\kappa,\mathbf{b}_T,\mu,\zeta)&=&\displaystyle 
  -\mathcal{P}_{i/k}^{[0]}(x,\kappa)
  \ln\left(\frac{\mu^2}{\mu_b^2}\right)
  +\mathcal{R}_{i/k}^{[1]}(x,\kappa)\\
\label{eq:finalres2} \\
\nonumber   &-&\displaystyle\delta_{ik}\delta(1-x)2C_i\left[\frac12\ln^2\left(\frac{\mu^2}{\mu_b^2}\right) -\left(K_i + \ln\left(\frac{\mu^2}{(1-\xi^2)\zeta}\right)\right)\ln\left(\frac{\mu^2}{\mu_b^2}\right) +\frac{\pi^2}{12}\right]\,,
\end{eqnarray}
where the rapidity divergence has cancelled, leaving an explicitly
finite result. We now just need to compute the residual functions
$\mathcal{R}_{i/k}^{[1]}$ to complete the picture.

\subsection{The unsubtracted GTMD
  correlators}\label{sec:unsubtractedGTMDs}

In this section, we present the results for the one-loop unsubtracted
parton-in-parton GTMD correlators $\Phi_{i/k}$ as defined in
Eq.~(\ref{eq:GTMDdefinition}). This calculation is functional to the
extraction of the residual functions $\mathcal{R}_{i/k}^{[1]}$ as well
as of the renormalisation constant $Z_{\Phi, i}$ in
Eq.~(\ref{eq:unsubtractedGTMDRenormalisation}).

Practically speaking, the computation is closely related to that of
the splitting functions $\mathcal{P}_{i/k}^{[0]}$ presented in
Ref.~\cite{Bertone:2022frx}. As already discussed in
Sect.~\ref{sec:matching}, the real contribution to the unsubtracted
parton-in-parton GTMD $\Phi_{i/k}^{[1],\rm real}$ is a simple
generalisation of the real contribution to the respective GPD
$F_{i/k}^{[1],\rm real}$ (see Eq.~(\ref{eq:GTMDreal})). In addition,
as again discussed in Sect.~\ref{sec:matching}, the virtual
contribution $\Phi_{i/k}^{[1],\rm virt}$ coincides with
$F_{i/k}^{[1],\rm virt}$. The only new element
w.r.t. Ref.~\cite{Bertone:2022frx} is that we need to retain, not only
the pole part of the correlator, but also the first finite correction
in the expansion in powers of $\epsilon$. This is precisely where the
residual functions $\mathcal{R}_{i/k}^{[1]}$ reside.

The main difference between the theoretical setup of this work and
that of Ref.~\cite{Bertone:2022frx} is the choice of the gauge. While
here we opted for the Lorentz gauge, in Ref.~\cite{Bertone:2022frx}
the light-cone gauge was used. However, a simple analysis shows that
this difference is immaterial, at least at one-loop accuracy. The
reasoning runs as follows. The GPD correlators are by construction
gauge invariant, signifying that the light-cone-gauge result of
Ref.~\cite{Bertone:2022frx} must coincide with a calculation in Lorenz
gauge.\footnote{I have explicitly verified this statement in the $q/q$
  channel by re-performing the calculation of
  Ref.~\cite{Bertone:2022frx} in Lorentz gauge.} Now, the differences
between the GPD and the GTMD correlators are:
\begin{enumerate}
\item a transverse displacement of the operator that causes the
  appearance of a phase in the real diagrams (see
  Sect.~\ref{sec:matching}),
\item the appearance of a transverse Wilson line at light-cone
  infinity.
\end{enumerate}
We have already discussed in Sect.~\ref{sec:matching} how to take care
of the phase while, in Lorentz gauge, the transverse Wilson line gives
no contribution. The conclusion is that, at least at one loop, the
light-cone-gauge calculation of the GPD correlators can be entirely
``recycled'' to obtain the GTMD correlators in Lorentz gauge. Based on
this conclusion, we refer the reader to Ref.~\cite{Bertone:2022frx}
for the details of the calculation. Here, we only report the result
for the bare unsubtracted GTMD correlators:
\begin{eqnarray}
\nonumber  \displaystyle  \hat{\Phi}_{i/k}^{[1]}(x,\xi,\mathbf{b}_T,\mu,\delta) &=&\displaystyle
                                                                           D_k(\xi)\bigg[-\frac{S_\epsilon}{\epsilon_{\rm
                                                                           IR}}\mathcal{P}_{i/k}^{[0],
                                                                           \rm
                                                                           real}(x,\kappa,\delta)-\mathcal{P}_{i/k}^{[0],
                                                                           \rm real}(x,\kappa,\delta) \ln
                                                                           \frac{\mu^2}{\mu_b^2}+
                                                                           \mathcal{R}_{i/k}^{[1]}(x,\kappa)\\
\label{eq:BareUnsubGTMD}\\
\nonumber &+&\displaystyle
    \delta_{ik}\delta(1-x) 2C_i\left(K_i-\ln(1-\xi^2)+2\ln\delta\right)\frac{\mu^{2\epsilon}S_{\epsilon}}{\epsilon_{\rm
    UV}}+\mathcal{O}(\epsilon)\bigg]\,,
\end{eqnarray}
where we have labelled the poles as IR or UV. Using the leading-order
result:
\begin{equation}
\hat{\Phi}_{i/k}^{[0]}(x,\xi,\mathbf{b}_T,\mu,\delta) = D_k(\xi) \delta_{ik}\delta(1-x)\,,
\end{equation}
that descend from Eq.~(\ref{eq:LOunsubtractedGTMDs}), we can extract
the $\overline{\mbox{MS}}$ renormalisation constant $Z_{\Phi,i}$ as
defined in Eq.~(\ref{eq:unsubtractedGTMDRenormalisation}) up to one
loop:
\begin{equation}
  Z_{\Phi,i}(\xi,\mu,\delta,\epsilon) = 1+a_s2C_i\left(K_i-\ln(1-\xi^2)+2\ln\delta\right)\frac{S_{\epsilon}}{\epsilon}+\mathcal{O}(a_s^2)\,.
\label{eq:ZBeamFunction}
\end{equation}

The residual functions $\mathcal{R}_{i/k}^{[1]}$ in
Eq.~(\ref{eq:BareUnsubGTMD}) are best given for non-singlet and
singlet GTMD combinations, respectively defined as:
\begin{equation}\label{eq:totalvalence}
  \Phi^- = \sum_{q}\Phi_{q/k}-\Phi_{\overline{q}/k}\,,\qquad
  \Phi^+=
\begin{pmatrix}
  \displaystyle \sum_{q}\Phi_{q/k}+\Phi_{\overline{q}/k}\\
  \Phi_{g/k}
\end{pmatrix}\,.
\end{equation}
The sums run over the active quark flavours and the anti-quark
correlators are defined as:
\begin{equation}
\Phi_{\overline{q}/k}(x,\dots) = - \Phi_{q/k}(-x,\dots)\,.
\end{equation}
In addition, it is convenient to introduce the following
decomposition:
\begin{equation}
\mathcal{R}^{\pm,[1]}(y,\kappa)=\theta(1-y)\mathcal{R}_1^{\pm,[1]}(y,\kappa)+\theta(\kappa-1)\mathcal{R}_2^{\pm,[1]}(y,\kappa)\,.
\label{eq:Rdecomposition}
\end{equation}
In the non-singlet sector one finds:
\begin{equation}
\left\{
\begin{array}{rcl}
\displaystyle \mathcal{R}_1^{-,[1]}(y,\kappa) &=& \displaystyle
                                        2C_F\frac{1-y}{1-\kappa^2y^2}\,,\\
\\
\displaystyle \mathcal{R}_2^{-,[1]}(y,\kappa) &=& \displaystyle 2 C_F \frac{(1-\kappa)y}{1-\kappa^2y^2}\,,
\end{array}
\right.
\label{eq:NonSingletResidualFuncs}
\end{equation}
while the singlet results are:
\begin{equation}
\begin{array}{ll}
\hspace{-15pt}\left\{
\begin{array}l
\displaystyle \mathcal{R}_{1,qq}^{+,[1]}(y,\kappa) = \displaystyle \mathcal{R}_1^{-,[1]}(y,\kappa)\,,\\
\\
\displaystyle \mathcal{R}_{2,qq}^{+,[1]}(y,\kappa) = \displaystyle 2
                                             C_F\frac{1-\kappa}{\kappa (1-\kappa^2y^2)}\,,
\end{array}
\right.
& 
\hspace{-15pt}\left\{
\begin{array}l
\displaystyle \mathcal{R}_{1,qg}^{+,[1]}(y,\kappa) = \displaystyle
                                             4n_fT_R\frac{y(1-y)}{(1-\kappa^2
                                             y^2)^2}\,,\\
\\
\displaystyle \mathcal{R}_{2,qg}^{+,[1]}(y,\kappa) = \displaystyle 4n_fT_R\frac{(1-\kappa)y^2}{(1-\kappa^2
                                             y^2)^2}\,,
\end{array}
\right.
\\
\\
\hspace{-15pt}\left\{
\begin{array}l
\displaystyle \mathcal{R}_{1,gq}^{+,[1]}(y,\kappa) = \displaystyle 2 C_F \frac{(1-\kappa^2)y}{1-\kappa^2y^2}\,,\\
\\
\displaystyle \mathcal{R}_{2,gq}^{+,[1]}(y,\kappa) = \displaystyle -2C_F \frac{1-\kappa^2}{\kappa(1-\kappa^2y^2)}\,,
\end{array}
\right.
&
\hspace{-15pt}\left\{
\begin{array}l
\displaystyle \mathcal{R}_{1,gg}^{+,[1]}(y,\kappa) = \displaystyle 8 C_A\frac{\kappa^2y(1-y)}{(1-\kappa^2y^2)^2}\,,\\
\\
\displaystyle \mathcal{R}_{2,gg}^{+,[1]}(y,\kappa) = \displaystyle C_A\frac{(1-\kappa)(1+\kappa-(1-7\kappa) \kappa^2 y^2)}{\kappa(1-\kappa^2y^2)^2}\,.
\end{array}
\right.
\end{array}
\label{eq:SingletResidualFuncs}
\end{equation}

All the expressions for $\mathcal{R}_1$ and $\mathcal{R}_2$ above are
singular at $y=1/\kappa$. It turns out that this singularity cancels
out in the combination in Eq.~(\ref{eq:Rdecomposition}) for
$\mathcal{R}^{-,[1]}$, $\mathcal{R}_{qq}^{+,[1]}$, and
$\mathcal{R}_{gq}^{+,[1]}$. Unfortunately this does not happen for
$\mathcal{R}_{qg}^{+,[1]}$ and
$\mathcal{R}_{gg}^{+,[1]}$. Specifically, for $\kappa>1$ and $y<1$,
one is left with undefined integrals of this kind:
\begin{equation}
  J=\int_x^1\frac {dy}{1-\kappa y} f(y)\,.
\end{equation}
Nevertheless, appropriately interpreting these integrals as
principal-values, one can use a trick presented in
Ref.~\cite{Bertone:2022frx} to obtain a numerically amenable
representation that reads:
\begin{equation}
  J = \int_x^1\frac {dy}{1-\kappa y}
  \left[f(y)-f\left(\frac{1}{\kappa}\right)\left(1+\theta\left(\kappa y-1\right)\frac{1-\kappa
        y}{\kappa
        y}\right)\right]+f\left(\frac{1}{\kappa}\right)\frac{1}{\kappa}\ln\left[\frac{\kappa(1-\kappa
      x)}{\kappa-1}\right]\,.
\label{eq:principalvaluexi}
\end{equation}
However, these integrals are clearly divergent for
$\kappa\rightarrow 1^+$ because in this limit the pole of the
integrand becomes an end-point singularity. As a consequence, as we
will explicitly see in Sect.~\ref{sec:numerics}, the GTMD correlator
will manifest a divergent behaviour around $x=\xi$. We also notice
that these divergences, being limited to $\mathcal{R}_{qg}^{+,[1]}$
and $\mathcal{R}_{gg}^{+,[1]}$, emerge from the convolution of the
matching functions with the gluon GPDs. Therefore, one may argue that
vanishing gluon GPDs at $x=\xi$ would guarantee the finiteness of the
GTMD correlator at $x=\xi$. However, this constraint on the gluon GPDs
seems to be hard to fulfil in full generality in that it should hold
at all scales.

In this respect, I can foresee two possible scenarios. In the first
scenario, the gluon GPDs obey a specific functional constraint such
that they remain null at $x=\xi$ at all scales. Incidentally, this
constraint seems to be possible to realise by means of the so-called
``shadow'' GPDs recently introduced in Ref.~\cite{Bertone:2021yyz}. In
the second scenario, the gluon GPDs are different from zero at $x=\xi$
causing the GTMD correlator to diverge in this point. To the best of
my knowledge, while even a discontinuity at $x=\xi$ at the level of
GPDs would cause the breaking of leading-twist collinear factorisation
in deeply-virtual Compton scattering (DVCS)~\cite{Diehl:2003ny}, a
divergent GTMD correlator at $x=\xi$ is not in principle forbidden. It
is interesting to observe that, beyond leading twist, discontinuities
at $x=\xi$ in DVCS partonic amplitudes do appear even in the collinear
case, see \textit{e.g.} Ref.~\cite{Kivel:2000rb, Kivel:2000cn,
  Radyushkin:2000jy}. However, in those papers it has been shown that
these discontinuities are such that they do not break factorisation
within the relevant accuracy.\footnote{I thank B. Pasquini for having
  drawn my attention to these works.}

As discussed in Ref.~\cite{Altinoluk:2012nt} in the context of DVCS,
it can also be argued that the singularity at $x=\xi$ is a consequence
of a kinematic enhancement due to the emission of soft-collinear
gluons that spoil the perturbative convergence in this
region. Analogously to the case of soft gluons in the threshold region
$x\rightarrow 1$ in deep-inelastic scattering~\cite{Catani:1989ne,
  Catani:1990rp}, their emission can be resummed to all orders in the
strong coupling $\alpha_s$ producing better-behaved results around
$x=\xi$~\cite{Altinoluk:2012nt}.\footnote{I am grateful to C. Mezrag
  for pointing me to Ref.~\cite{Altinoluk:2012nt}.} In any event, this
point deserves further investigations.

\subsection{Forward limit of the matching functions}\label{sec:forward_limit}

To the best of my knowledge, this is the first calculation of GTMD
matching functions to one-loop accuracy ever performed. As such, there
are no previous calculations that can be used to compare these results
to. However, these matching functions need to tend to their well-known
forward counterpart used to match TMDs onto collinear PDFs.

To perform this check, we first observe that the structure of
Eq.~(\ref{eq:finalres2}) is the same as that found in the TMD case:
see for instance Eq.~(4.8) of Ref.~\cite{Echevarria:2011epo}. The
forward limit is realised by taking the limit $\xi\rightarrow 0$ that
is equivalent to $\kappa\rightarrow 0$. In this limit, we have already
verified in Ref.~\cite{Bertone:2022frx} that the GPD splitting
functions $\mathcal{P}_{i/k}^{[0]}$ tend to the Altarelli-Parisi
splitting functions. We can therefore safely simplify
Eq.~(\ref{eq:finalres2}) by setting $\mu=\mu_b$ in such a way that all
logarithms vanish. Expressing the matching functions in the basis
defined in Eq.~(\ref{eq:totalvalence}), we find:
\begin{equation}
\begin{array}{rcl}
  \mathcal{C}^{[1],-}(y,\kappa,\mathbf{b}_T,\mu_b,\zeta) &=&\displaystyle
  \mathcal{R}^{[1],-}(y,\kappa)-C_q\frac{\pi^2}{6}\delta(1-y)\,,\\
\\
  \mathcal{C}_{ik}^{[1],+}(y,\kappa,\mathbf{b}_T,\mu_b,\zeta) &=&\displaystyle
  \mathcal{R}_{ik}^{[1],+}(y,\kappa)-C_i\frac{\pi^2}{6}\delta_{ik}\delta(1-y)\,.
\end{array}
\end{equation}
When taking the limit $\kappa\rightarrow 0$, the term proportional to
$\theta (\kappa-1)$ in the decomposition of $\mathcal{R}^{[1],\pm}$ in
Eq.~(\ref{eq:Rdecomposition}) drops. This leaves only the term
proportional to $\theta(1-y)$ that, introduced in the convolution
integral as defined in Eq.~(\ref{eq:matchingOnToGPDs}), turns it into
a standard Mellin convolution. Therefore, all we need to do is to take
the limit $\kappa\rightarrow 0$ of the functions
$\mathcal{R}_1^{[1],\pm}$. This is easily done using
Eqs.~(\ref{eq:NonSingletResidualFuncs})
and~(\ref{eq:SingletResidualFuncs}) and the result is:
\begin{equation}
\begin{array}{rcl}
\displaystyle \lim_{\kappa\rightarrow
  0}\mathcal{C}^{[1],-}(y,\kappa,\mathbf{b}_T,\mu_b,\zeta) =   \lim_{\kappa\rightarrow
  0}\mathcal{C}_{qq}^{[1],+}(y,\kappa,\mathbf{b}_T,\mu_b,\zeta)
  &=&\displaystyle  2C_F(1-y)-C_F\frac{\pi^2}{6}\delta(1-y)\,,\\
\\
\displaystyle\lim_{\kappa\rightarrow
  0}\mathcal{C}_{qg}^{[1],+}(y,\kappa,\mathbf{b}_T,\mu_b,\zeta)
  &=&\displaystyle  4n_f T_R y(1-y)\,,
\\
\\
\displaystyle\lim_{\kappa\rightarrow
  0}\mathcal{C}_{gq}^{[1],+}(y,\kappa,\mathbf{b}_T,\mu_b,\zeta)
  &=&\displaystyle  2C_Fy\,,
\\
\\
\displaystyle\lim_{\kappa\rightarrow
  0}\mathcal{C}_{gg}^{[1],+}(y,\kappa,\mathbf{b}_T,\mu_b,\zeta)
  &=&\displaystyle  -C_A\frac{\pi^2}{6}\delta(1-y)\,,
\end{array}
\end{equation}
that indeed coincide with the TMD results (see, \textit{e.g.},
Ref.~\cite{Echevarria:2016scs}).

\subsection{Anomalous dimensions}\label{sec:anomalous_dimensions}

As discussed in Sect.~\ref{sec:renormalisation_and_evolution}, the
knowledge of the renormalisation constants of soft function and
unsubtracted GTMD correlators allows us to extract the anomalous
dimensions that govern the evolution of the normalised GTMD
correlators (see Eq.~(\ref{eq:evolutionequations})). Using
Eq.~(\ref{eq:evolutionkernel}) with the normalisation constants
$Z_{S,i}$ and $Z_{\Phi,i}$ respectively given in
Eqs.~(\ref{eq:ZSoftFunction}) and~(\ref{eq:ZBeamFunction}), such that:
\begin{equation}
{Z}_{S,i}^{1/2}Z_{\Phi,i}^{-1} =1 - a_s
 2C_i\left[\frac{S_\epsilon^2}{\epsilon^2}+\frac{S_\epsilon}{\epsilon}\left(K_i+\ln\left(\frac{\mu^2}{(1-\xi^2)Q^2}\right)\right)+\ln\left(\frac{\mu^2}{\mu_b^2}\right)\ln\left(\frac{\zeta}{Q^2}\right)\right]+\mathcal{O}(a_s^2)\,,
\end{equation}
one immediately finds:
\begin{equation}
\begin{array}{rcl}
\displaystyle K_i(\mathbf{b}_T,\mu) &=&\displaystyle 
-a_s4C_i\ln\left(\frac{\mu^2}{\mu_b^2}\right)+\mathcal{O}(a_s^2)\,,\\
\\
\displaystyle \gamma_i(\mu,\zeta)  &=&\displaystyle  a_s
 4C_i\left(K_i+\ln\left(\frac{\mu^2}{(1-\xi^2)\zeta}\right)\right)
 +\mathcal{O}(a_s^2)\,.
\end{array}
\label{eq:AnomDimsOneLoop}
\end{equation}
Notice that, for computing the total derivative of
${Z}_{S,i}^{1/2}Z_{\Phi,i}^{-1}$ w.r.t. $\mu$ as prescribed by the
second equation in Eq.~(\ref{eq:evolutionkernel}), we have used the
identity:
\begin{equation}
\frac{d \ln[{Z}_{S,i}^{1/2}Z_{\Phi,i}^{-1}]}{d\ln\mu} = \frac{\partial \ln[{Z}_{S,i}^{1/2}Z_{\Phi,i}^{-1}]}{\partial \ln\mu}+2(-\epsilon a_s+\beta(a_s)) \frac{\partial \ln[{Z}_{S,i}^{1/2}Z_{\Phi,i}^{-1}]}{\partial a_s}\,,
\end{equation}
where $\beta(a_s)=\mathcal{O}(a_s^2)$ is the four-dimensional QCD
$\beta$ -function that governs the running of the strong coupling:
\begin{equation}
  \frac{da_s}{d\ln\mu^2}=\beta(a_s(\mu))\,.
\end{equation}
The cusp anomalous dimension is extracted using
Eq.~(\ref{eq:TMDCrossDerivative}) and equals:\footnote{Notice that
  deriving either $K_i$ w.r.t. $\mu$ or $\gamma_i$ w.r.t. $\zeta$
  consistently gives the same result.}
\begin{equation}
  \gamma_{K,i}(a_s)  = a_s8C_i + \mathcal{O}(a_s^2)\,.
\label{eq:CuspOneLoop}
\end{equation}
Finally, Eqs.~(\ref{eq:AnomDimsOneLoop}) and~(\ref{eq:CuspOneLoop})
allow us to extract the leading-order coefficients:
\begin{equation}
\begin{array}{l}
K_i^{[0]} = 0\,,\\
\\
\gamma_{F,i}^{[0]} = 4C_iK_i\,,\\
\\
\gamma_{K,i}^{[0]} = 8C_i\,.
\end{array}
\end{equation}
These results coincide with those obtained in the TMD framework (see,
\textit{e.g.}, Ref.~\cite{Collins:2017oxh} and references therein for
a recent overview). It is therefore legitimate to expect that the same
holds true to all orders such that we can use the known higher-order
corrections to these quantities to evolve GTMDs to higher
accuracies. We will do so in the next section.

\section{Numerical setup}\label{sec:numerics}

In this section, we present numerical results showing how the matching
functions computed above can be used to reconstruct realistic GTMDs by
combining: a sensible model for GPDs along with collinear off-forward
evolution, perturbative TMD evolution, and a model for the
non-perturbative transverse effects as determined by recent TMD
extractions.

The limitation of this procedure is that it can only be applied to
GTMDs that have, at the same time, a projection on a GPD \textit{and}
on a TMD.\footnote{By projection of a GTMD on a GPD, it is meant the
  integral of the former over the partonic transverse momentum
  $\mathbf{k}_T$ or, alternatively (and perhaps more correctly), the
  GPD on which the GTMD matches at large $\mathbf{k}_T$. While the
  projection of a GTMD on a TMD is its forward limit,
  $\Delta\rightarrow 0$. It is well-known that several GTMDs do not
  have projections either on GPDs or on TMDs, in the sense that their
  contribution to the correlator vanishes by
  kinematics~\cite{Meissner:2009ww, Lorce:2013pza}.} This largely
restricts the range of possibilities. In addition, to give an estimate
of such GTMDs, it is highly desirable that the GPDs and TMDs on which
they project are, to some extent, phenomenologically known.

Using the results of Refs.~\cite{Meissner:2009ww,Lorce:2013pza}, the
renormalised GTMD correlator, Eq.~(\ref{eq:renormalisedGTMD}), has the
following tensorial decomposition:
\begin{equation}
  \mathcal{F}_{i/H} =
  \frac1{2M}\overline{u}(P_{\rm
    out})\left[F_{1,1}^{i}+\frac{i\sigma^{\mathbf{k}_Tn}}{n\cdot P}F_{1,2}^{i}+\frac{i\sigma^{\mathbf{\Delta}_Tn}}{n\cdot P}F_{1,3}^{i}+\frac{i\sigma^{\mathbf{k}_T\mathbf{\Delta}_T}}{M^2}F_{1,4}^{i}\right]u(P_{\rm in})\,,
\end{equation}
with $u$ and $\overline{u}$ being respectively the spinors of the
incoming and outgoing hadron $H$ having mass $M$, and where we have
used the shorthand notation
$a_\mu b_\nu\sigma^{\mu\nu}\equiv \sigma^{ab}$. The scalar
coefficients $F_{1,l}$, with $l=1,\dots,4$, are the actual GTMD
distributions (GTMDs for short). Each of them can be split into a
T-even and a T-odd part, both real, as follows:
\begin{equation}
  F_{1,l}^{i}=F_{1,l}^{i,e}+i F_{1,l}^{i,o}\,.
\end{equation}
The peculiarity of $F_{1,l}^{i,e}(F_{1,l}^{i,o})$ is that it conserves
(flips) the sign upon swapping of the Wilson line from past to future
pointing and viceversa.

The one GTMD that fulfils all the requirements stated above is
$F_{1,1}^{i,e}$. As a matter of fact, in $\mathbf{b}_T$ space and at
small $\mathbf{b}_T$, $F_{1,1}^{i,e}$ matches on the following
combination of GPDs:
\begin{equation}
  F_{1,1}^{i,e}(x,\xi,b_T,t,\mu,\zeta) = \mathcal{C}_{i/j}(x,\kappa,
  b_T,\mu,\zeta)\mathop{\otimes}_{x}\left[(1-\xi^2)H_{j}(x,\xi,t,\mu)-\xi^2
    E_{j}(x,\xi,t,\mu)\right]\,,
\label{eq:generalscalematching}
\end{equation}
where we have replaced $\mathbf{b}_T$ with $b_T$ everywhere because
$F_{1,1}^{i,e}$ does not depend of the direction on
$\mathbf{b}_T$. Currently, different models for $H_j$ and $E_j$
exist. In addition, the forward limit of $F_{1,1}^{i,e}$ is the fully
unpolarised TMD $f_{1,i}$, that is:
\begin{equation}
  \lim_{\xi,t\rightarrow 0} F_{1,1}^{i,e}(x,\xi, b_T,t,\mu,\zeta) = f_{1,i}(x, b_T,\mu,\zeta)\,.
\end{equation}
Accurate phenomenological determinations from data of $f_{1,i}$, with
$i=q$, have recently been presented.

In order to compute $F_{1,1}^{i,e}$ numerically, we employ a standard
procedure used in TMD factorisation (see, \textit{e.g.},
Ref.~\cite{Bacchetta:2019sam}). Let us first state the complete
formula and then comment on it:
\begin{eqnarray}
\nonumber F_{1,1}^{i,e}(x,\xi,b_T,t,\mu,\zeta) &=&\displaystyle
                                           \mathcal{C}_{i/j}(x,\kappa,
                                           b_*,\mu_{b_*},\mu_{b_*}^2)\mathop{\otimes}_{x}\left[(1-\xi^2)H_{j}(x,\xi,t,\mu_{b_*})-\xi^2
                                           E_{j}(x,\xi,t,\mu_{b_*})\right]\\
\nonumber  \\
\label{eq:bspaceexpression}  &\times&R_i\left[(\mu,\zeta)\leftarrow(\mu_{b_*},\mu_{b_*}^2)\right]\\
\nonumber\\
\nonumber&\times& \displaystyle f_{\rm NP}(x,b_T,(1-\xi^2)\zeta)\,.
\end{eqnarray}
First, we have introduced the so-called $b_*$-prescription with this
particular functional form~\cite{Bacchetta:2017gcc, Bacchetta:2019sam,
  Bacchetta:2022awv}:
\begin{equation}
b_*\equiv
b_*(b_T)=\frac{b_0}{Q}\left(\frac{1-\exp\left(-\frac{b_T^4Q^4}{b_0^4}\right)}{1-\exp\left(-\frac{b_T^4}{b_0^4}\right)}\right)^{\frac14}\,,
\label{eq:bstar}
\end{equation}
and the associated scale $\mu_{b_*}=b_0/b_*(b_T)$. The scale $Q$ is
usually identified with the physical hard scale of the process under
consideration. In all cases, one must have
$Q\simeq \mu\simeq \sqrt{\zeta}$. Since the $\mathbf{k}_T$-space GTMD
is obtained through a Fourier transform of the $\mathbf{b}_T$-space
expression, the scope of the $b_*$-prescription is to prevent the
scale $\mu_{b_*}$ from becoming too low such to enter the
non-perturbative regime when $b_T\rightarrow \infty$. Indeed, if
$\mu_{b_*}$ became of the order of $\Lambda_{\rm QCD}$ or smaller, any
perturbative calculation would be invalidated. To this purpose,
Eq.~(\ref{eq:bstar}) is engineered in a way that:
\begin{equation}
  \lim_{b_T\rightarrow \infty}\mu_{b_*}(b_T)= 1\mbox{ GeV}.
\end{equation}
In addition, Eq.~(\ref{eq:bstar}) is also such that:
\begin{equation}
  \lim_{b_T\rightarrow 0}\mu_{b_*}(b_T)= Q.
\end{equation}
This last property is not strictly needed but it helps keep under
better control the large-$\mathbf{k}_T$ region~\cite{Bozzi:2005wk,
  Bizon:2018foh}.

While the $b_*$-prescription ensures the applicability of perturbation
theory, it remains necessary to keep into account non-perturbative
transverse effects. This is done through the function $f_{\rm NP}$ in
the third line of Eq.~(\ref{eq:bspaceexpression}) that parameterises
non-perturbative large-$b_T$ effects. See
Refs.~\cite{Bacchetta:2019sam, Scimemi:2019cmh, Bacchetta:2022awv} for
recent global determinations of $f_{\rm NP}$ from experimental data in
the TMD framework. For our numerical estimates, we will use the
determination of $f_{\rm NP}$ from Ref.~\cite{Bacchetta:2019sam},
henceforth referred to as PV19 (for Pavia 2019), for both quark and
gluon distributions.\footnote{Notice that the $f_{\rm NP}$ extracted
  in Ref.~\cite{Bacchetta:2019sam} strictly applies to quark
  distributions. Since, to the best of my knowledge, there is no
  comparably reliable analogue for the gluon, we will use the same
  $f_{\rm NP}$ also for this distribution.} $f_{\rm NP}$ in
Eq.~(\ref{eq:bspaceexpression}) only depends on $x$, $b_T$, and the
combination $(1-\xi^2)\zeta$ (the latter related to the evolution
pattern). However, in the GTMD case this function should in principle
also depend on $\xi$ and $t$. Since $f_{\rm NP}$ from PV19 is obtained
from a fit of TMDs where $\xi$ and $t$ are identically zero, we do not
have access to the non-perturbative transverse dependence on these
variables. Nonetheless, this dependence is expected to be mild in that
most of the effect is accounted for by the collinear
GPDs.\footnote{The argument behind this statement is the fact that
  $f_{\rm NP}$ is effectively obtained as a ratio of GTMDs computed a
  different values of $b_T$, see for example
  Ref.~\cite{Bacchetta:2019sam} for a discussion in the TMD
  framework. As a consequence, any dependence on kinematic variables
  is expected to be relatively mild. This turns out to be the case for
  the longitudinal momentum fraction $x$~\cite{Bacchetta:2019sam}.}

The first line of Eq.~(\ref{eq:bspaceexpression}) corresponds to the
matching onto the collinear GPDs as in
Eq.~(\ref{eq:generalscalematching}). However, the scales at which the
matching is performed are chosen to be $\mu=\sqrt{\zeta}=\mu_{b_*}$ so
that the matching functions are free of potentially large
logarithms. The GPDs $H_i$ and $E_i$ are computed using the
Goloskokov-Kroll (GK) model~\cite{Goloskokov:2005sd,
  Goloskokov:2007nt, Goloskokov:2009ia} at the initial scale
$\mu_0=2$~GeV and appropriately evolved to $\mu_{b_*}$ through
collinear evolution.

The evolution to the hard scales $\mu$ and $\zeta$ is provided by the
factor $R_i$ in the second line of Eq.~(\ref{eq:bspaceexpression}),
often referred to as Sudakov form factor. This factor results from the
solution of the evolution equations in
Eq.~(\ref{eq:evolutionequationsExp}) and reads:
\begin{equation}
\begin{array}{rcl}
  \displaystyle  R_i\left[(\mu,\zeta)\leftarrow
  (\mu_{b_*},\mu_{b_*}^2)\right] &=&\displaystyle \exp
                                     \bigg\{ K_i(b_*,\mu_{b_*}) \ln \frac{\sqrt{(1-\xi^2)\zeta}}{\mu_{b_*}} \\
  \\
                                 &+& \displaystyle
                                     \int_{\mu_{b_*}}^{\mu} \frac{d\mu'}{\mu'}\left[ \gamma_{F,i}
                                     (\alpha_s(\mu')) - \gamma_{K,i} (\alpha_s(\mu')) \ln
                                     \frac{\sqrt{(1-\xi^2)\zeta}}{\mu'} \right] \bigg\} \, .
\end{array}
\end{equation}
As discussed in Sect.~\ref{sec:renormalisation_and_evolution}, the
anomalous dimension $K_i(b_*,\mu_{b_*})$, $\gamma_{F,i}$, and
$\gamma_{K,i}$ are computable in perturbation theory and they (are
expected to) coincide with their TMD counterparts.

Finally, we can obtain the $F_{1,1}^{i,e}$ in $\mathbf{k}_T$ space by
taking a 2D Fourier transform of Eq.~(\ref{eq:bspaceexpression})
w.r.t. $\mathbf{b}_T$ that, with abuse of notation, gives:
\begin{equation}
F_{1,1}^{i,e}(x,\xi,k_T,t,\mu,\zeta)=\frac1{2\pi}\int_{0}^{\infty}db_T\,b_T
J_0(k_Tb_T) F_{1,1}^{i,e}(x,\xi,b_T,t,\mu,\zeta)\,,
\label{eq:kspaceexpression}
\end{equation}
where $J_0$ is a Bessel function of the first kind.

The theoretical precision of the formula in
Eq.~(\ref{eq:bspaceexpression}) is usually expressed in terms of
logarithmic accuracy. Tab.~1 of Ref.~\cite{Bacchetta:2019sam} tells us
that the knowledge of the matching functions $\mathcal{C}_{i/j}$ up to
$\mathcal{O}(\alpha_s)$ would in principle allow us to reach
next-to-next-to-leading logarithmic (NNLL) accuracy. In the case of
GTMDs, the only obstacle to reaching this accuracy is the evolution of
collinear GPDs that, as of today, is publicly available only up to
leading order~\cite{Bertone:2022frx}. However, in the following we
will use a NNLL-accurate setup, except for the GPD evolution for which
the $\mathcal{O}(\alpha_s)$ (\textit{i.e.} leading-order) splitting
functions are used. This practically means that $K_i$ and
$\gamma_{F,i}$ are computed at $\mathcal{O}(\alpha_s^2)$,
$\gamma_{K,i}$ at $\mathcal{O}(\alpha_s^3)$, $\mathcal{C}_{i/j}$ at
$\mathcal{O}(\alpha_s)$, and the evolution of the strong coupling
$\alpha_s$ is computed using a $\beta$ function truncated at
$\mathcal{O}(\alpha_s^3)$. In the following, in order to simplify the
presentation, we will set $\mu=\sqrt{\zeta}=Q$.

The numerical implementation of the GTMD in
Eq.~(\ref{eq:bspaceexpression}) and its Fourier transform
Eq.~(\ref{eq:kspaceexpression}) makes use of a compound of publicly
available tools. Specifically, the GK model for the GPDs $H_i$ and
$E_i$ at the initial scale $\mu_0$ is provided by {\tt
  PARTONS}~\cite{Berthou:2015oaw}\footnote{\url{https://partons.cea.fr/partons/doc/html/index.html}};
their collinear evolution as well as the perturbative ingredients
relevant to the matching and the Sudakov form factor are provided by
{\tt APFEL++}~\cite{Bertone:2013vaa,
  Bertone:2017gds}\footnote{\url{https://github.com/vbertone/apfelxx}};
the non-perturbative function $f_{\rm NP}$ and the Fourier transform
are instead provided by {\tt
  NangaParbat}~\cite{Bacchetta:2019sam}\footnote{\url{https://github.com/MapCollaboration/NangaParbat}}. The
resulting code is
public\footnote{\url{https://github.com/vbertone/GTMDMatchingFunctions}}
and can be used to reproduce the results shown below.

\begin{figure}[t]
  \centering
  \includegraphics[width=0.7\textwidth]{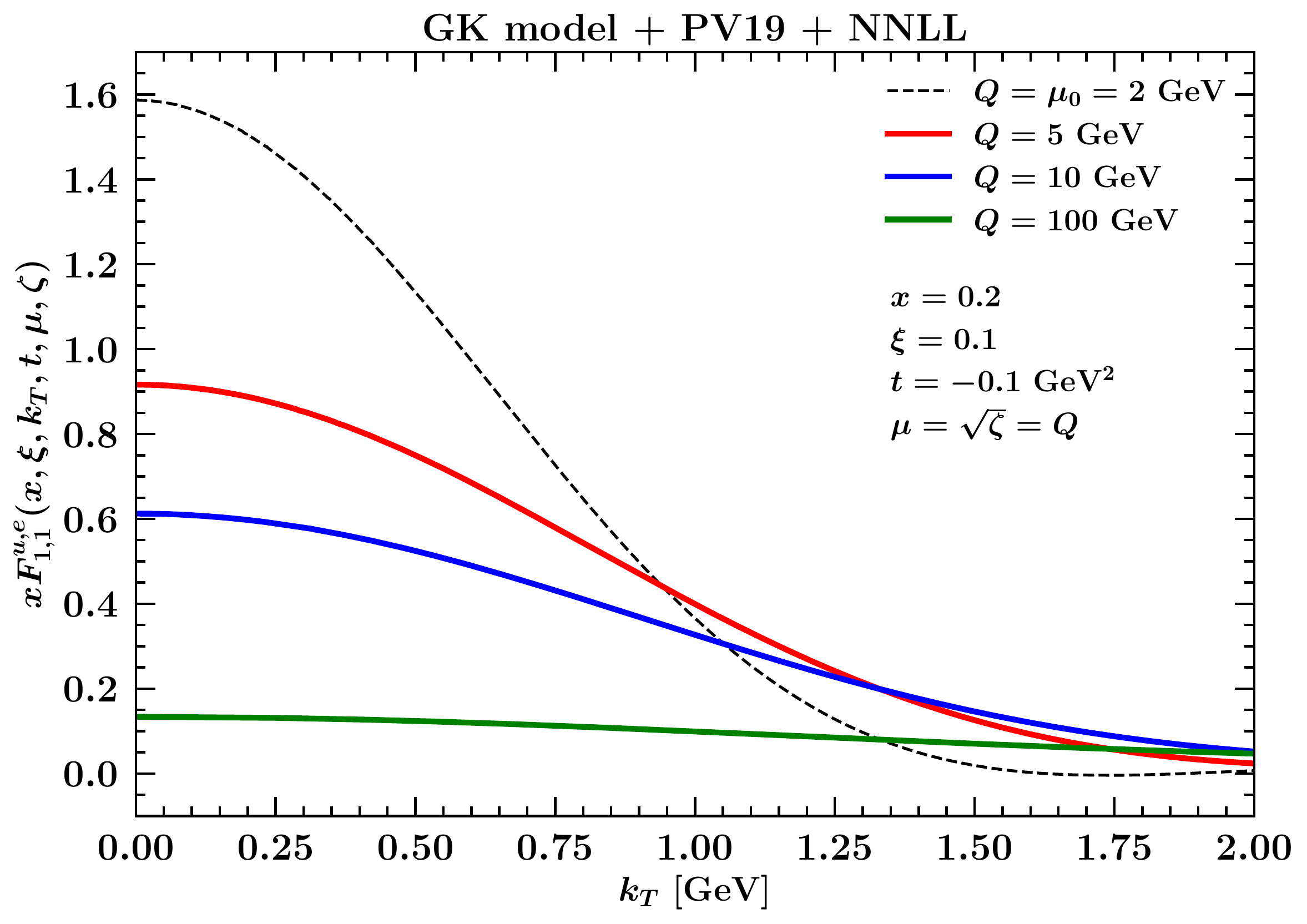}
  \vspace{20pt}
  \caption{$xF_{1,1}^{u,e}$ (up quark) vs. $k_T$ at $x=0.2$,
    $\xi=0.1$, and $t=-0.1$~GeV$^2$ for four different values of the
    hard scale $Q=\mu=\sqrt{\zeta}$: $Q=2$~GeV (blacked dashed curve,
    GPD-model at the initial scale $\mu_0$), $Q=5$~GeV (solid red
    curve), $Q=10$~GeV (solid blue curve), and $Q=100$~GeV (solid
    green curve). $F_{1,1}^{u,e}$ is computed at NNLL accuracy (see
    text) using the GK model for the GPDs and the PV19 determination
    for the non-perturbative transverse
    component. \label{fig:GTMDMatchingkTQu}}
\end{figure}
We can finally present a selection of quantitative results. We start
with Fig.~\ref{fig:GTMDMatchingkTQu} that shows the behaviour
w.r.t. $k_T$ of the up-quark GTMD $F_{1,1}^{u,e}$ at fixed values of
$x$, $\xi$, and $t$ for different values of the hard scale
$Q=\mu=\sqrt{\zeta}$. The curve at the GPD initial scale $Q=\mu_0$ is
also shown. The typical broadening of the $k_T$ distribution caused by
the Sudakov form factor as $Q$ increases is clearly
observed~\cite{Aybat:2011zv, Echevarria:2016mrc}. We also note that
the curve at $\mu_0$ becomes negative at large $k_T$ as a consequence
of the specific functional form $f_{\rm NP}$ used here. The gluon
distribution behaves very similarly, therefore we do not show the
corresponding plot.

\begin{figure}[t]
  \centering
  \includegraphics[width=0.7\textwidth]{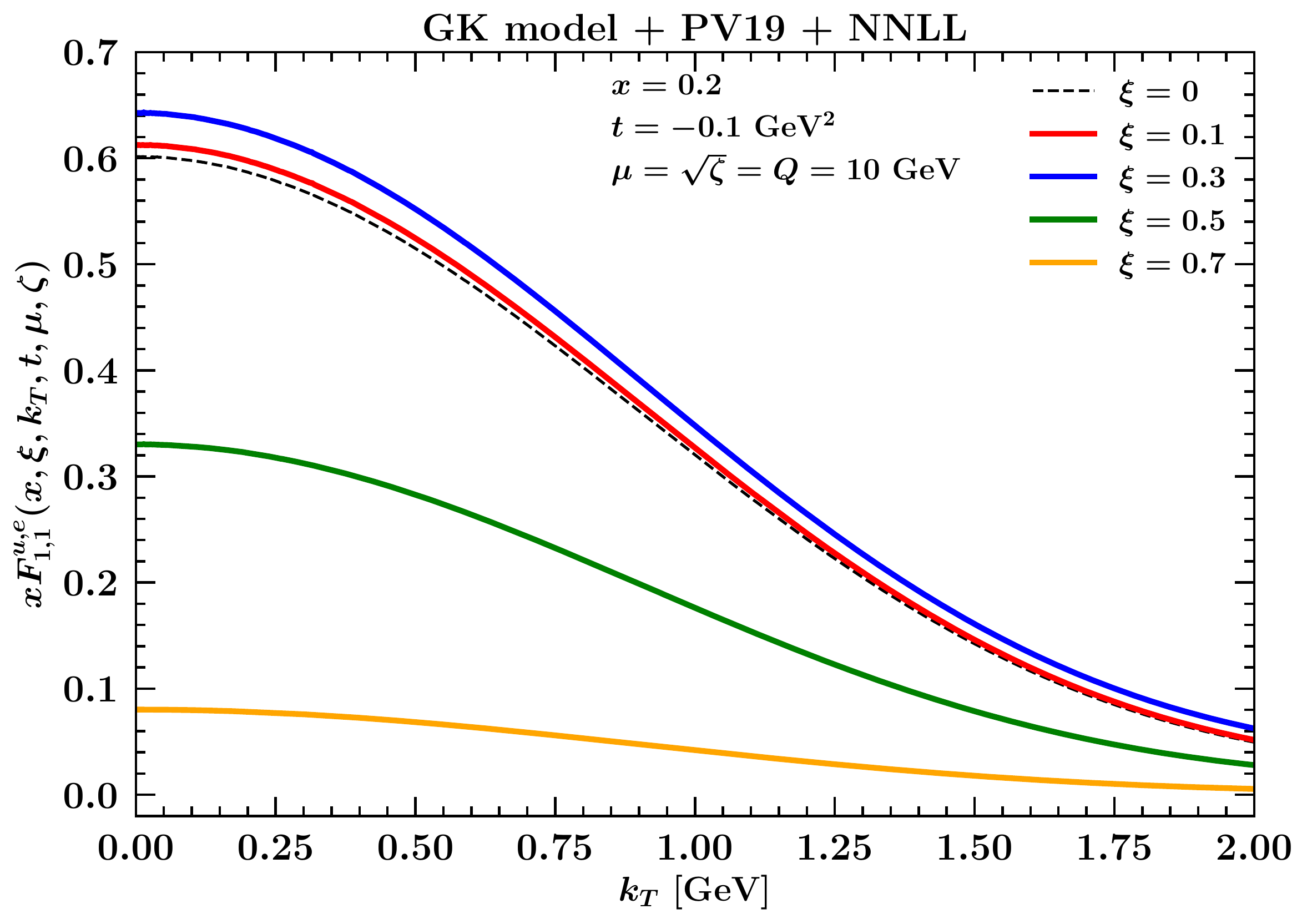}
  \vspace{20pt}
  \caption{$xF_{1,1}^{u,e}$ (up quark) vs. $k_T$ at $x=0.2$,
    $t=-0.1$~GeV$^2$, and $Q=\mu=\sqrt{\zeta}=10$~GeV for five
    different values of the skewness: $\xi=0$ (blacked dashed curve),
    $\xi=0.1$ (solid red curve), $\xi=0.3$ (solid blue curve),
    $\xi=0.5$ (solid green curve), and $\xi=0.7$ (solid orange
    curve). $F_{1,1}^{u,e}$ is computed at NNLL accuracy (see text)
    using the GK model for the GPDs and the PV19 determination for the
    non-perturbative transverse
    component.\label{fig:GTMDMatchingkTxiu}}
\end{figure}
It is also interesting to look at how the $k_T$ dependence of
$F_{1,1}^{i,e}$ changes with $\xi$. To this purpose,
Fig.~\ref{fig:GTMDMatchingkTxiu} shows curves for the up-quark
distribution at fixed values of $x$, $t$, and $Q$ for different values
of $\xi$. The black dashed line corresponds to $\xi=0$ that is, up to
a fully non-perturbative $t$ dependence of the underlying GPDs, the
``partonic'' forward limit. The behaviour in $\xi$ is clearly non
monotonic but exhibits a strong suppression as the value of $\xi$
approaches one. Again, the gluon distribution does not present any
further salient features and thus is not shown.

\begin{figure}[t]
  \centering
  \includegraphics[width=0.7\textwidth]{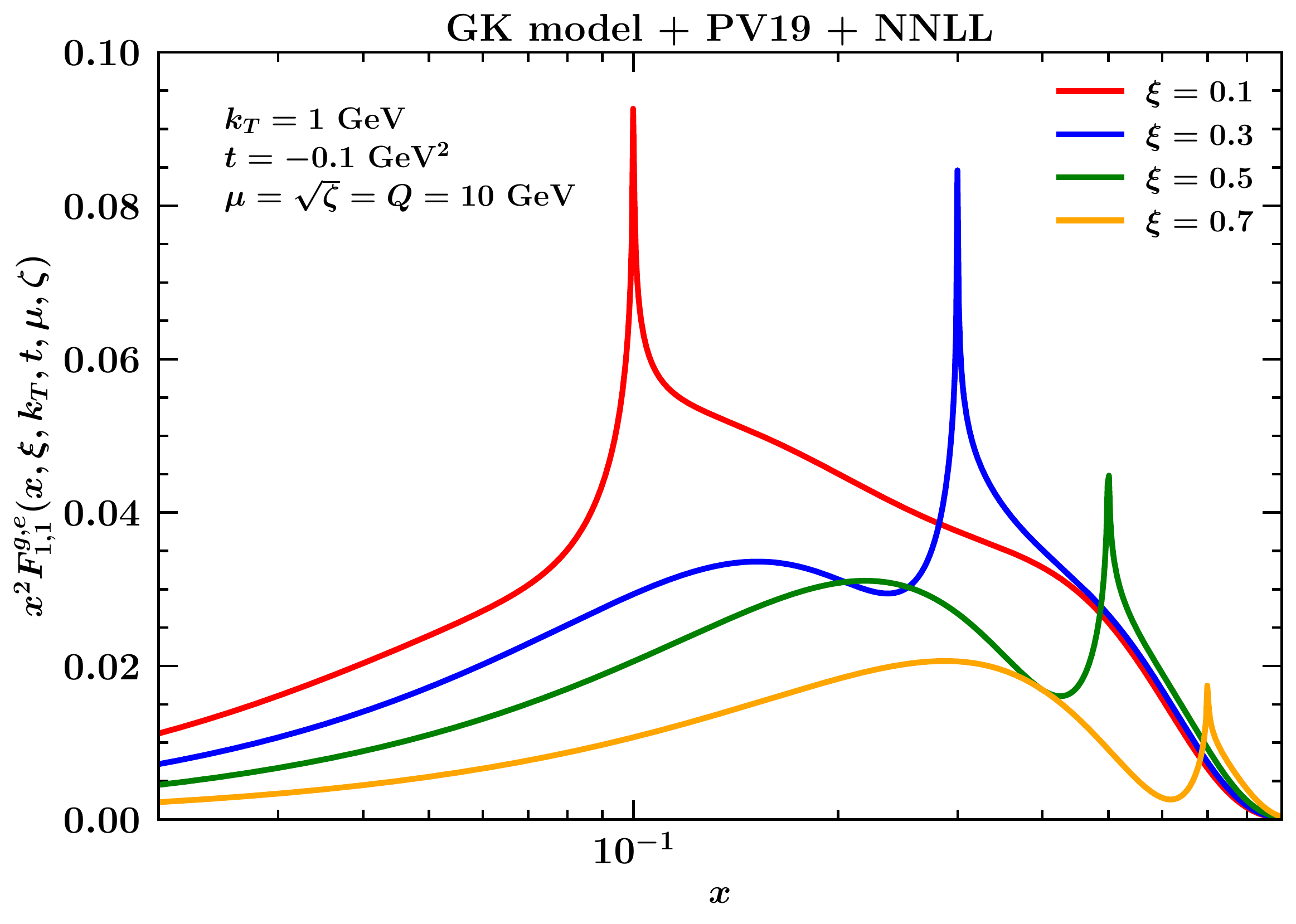}
  \vspace{20pt}
  \caption{$x^2F_{1,1}^{g,e}$ (gluon) vs. $x$ at $k_T=1$~GeV,
    $t=-0.1$~GeV$^2$, and $Q=\mu=\sqrt{\zeta}=10$~GeV for four
    different values of the skewness: $\xi = 0.1$ (red curve),
    $\xi = 0.3$ (blue curve), $\xi = 0.5$ (green curve), and
    $\xi = 0.7$ (orange curve). $F_{1,1}^{g,e}$ is computed at NNLL
    accuracy (see text) using the GK model for the GPDs and the PV19
    determination for the non-perturbative transverse
    component.\label{fig:GTMDMatchingxxig}}
\end{figure}
Finally, we consider the behaviour of $F_{1,1}^{i,e}$ as a function of
the longitudinal momentum fraction $x$. As pointed out in
Sect.~\ref{sec:unsubtractedGTMDs}, $\mathcal{R}_{qg}^{+,[1]}$ and
$\mathcal{R}_{gg}^{+,[1]}$ cause a divergence at $x=\xi$ in both quark
and gluon GTMD correlators. To study this effect, we concentrate on
the gluon GTMD $F_{1,1}^{g,e}$ and in Fig.~\ref{fig:GTMDMatchingxxig}
we show this distribution (weighted by $x^2$ to improve the
readability of the plot) as a function of $x$ at fixed values of
$k_T$, $t$, and $Q$ for different values of $\xi$. It is evident that
indeed the curves exhibit a divergence at $x=\xi$ that is a
manifestation of the divergence of the principal-valued integral in
Eq.~(\ref{eq:principalvaluexi}) for $\kappa\rightarrow 1^+$. The
conclusion is that GTMDs obtained through perturbative matching onto
GPDs beyond tree level are undefined at $x=\xi$. Whether this is an
acceptable feature of GTMDs remains an open question. As already
mentioned in Sect.~\ref{sec:unsubtractedGTMDs}, a possibility to
remove the divergence is to require the gluon GPDs to vanish at
$x=\xi$ at all scales.

\section{Conclusions}\label{sec:conclusions}

The main result of this paper is the calculation of the complete set
of one-loop unpolarised GTMD matching functions. These perturbative
functions allow one to express GTMDs at large partonic transverse
momentum $\mathbf{k}_T$ (or equivalently at low $\mathbf{b}_T$) in
terms of GPDs. They are obtained employing a rapidity-divergence-free
definition of the GTMD correlator~\cite{Echevarria:2016mrc} that
combines the soft function with the unsubtracted GTMD correlator. In
order to carry out the calculation, the principal-value regulator for
rapidity divergences has been used, allowing us to obtain one-loop
results for soft function, confirming previous results, and
unsubtracted GTMD correlator, that are instead original. By
renormalising the UV divergences of these objects, we have also
extracted the one-loop anomalous dimensions, again confirming known
results. In addition, as a consistency check, we have verified that in
the limit $\xi\rightarrow 0$ the one-loop GTMD matching functions thus
obtained reproduce the well-know TMD results.

In the last part of the paper, we have presented a quantitative study
by implementing the matching functions and combining them with other
perturbative and non-perturbative ingredients necessary to fully
reconstruct the GTMDs $F_{1,1}^{i,e}$, with $i=q,g$, at NNLL
accuracy. We have studied their $\mathbf{k}_T$ dependence along with
the scale evolution and the dependence on the skewness $\xi$. An
interesting consequence of our calculation is that, beyond tree level,
GTMD matching functions yield GTMDs that exhibit a pole at
$x=\xi$. The origin of this singularity is unclear and may signal the
need of additional theoretical constraints on the underlying gluon
GPDs. A more extensive investigation on this point is left for a
future study. Finally, we stress that the numerical results presented
here are readily accessible in that all of the relevant ingredients
are available in public codes. A code that consistently combines them
to reproduce the plots of this paper is also released with the paper.

The calculation presented here is limited to the unpolarised twist-2
Lorentz structure (see Eq.~(\ref{eq:GTMDdefinition})). The extension
of this study to the remaining twist-2 structures is planned for the
future. This will require computing the corresponding splitting
functions along the lines of Ref.~\cite{Bertone:2022frx} as well as
the one-loop corrections to the unsubtracted GTMD
correlators. Phenomenological applications of the GTMDs presented here
are also envisaged. Specifically, the computation of measurable (and
possible measured) cross sections using the GTMDs presented in this
paper would allow us to gauge the accuracy of the calculation as well
as the reliability of the underlying non-perturbative ingredients,
such as the GPDs and the TMDs.

\section*{Acknowledgements}

I acknowledge stimulating discussions with S.~Bhattacharya,
H.~Dutrieux, M.~G.~Echevarria, and H.~Moutarde. I am particularly
grateful to C.~Mezrag B.~Pasquini, and S.~Rodini for a critical
reading of the manuscript. This work was supported by the European
Union’s Horizon 2020 research and innovation programme under grant
agreement STRONG 2020 - No 824093.

\newpage

\bibliographystyle{ieeetr}
\bibliography{Bibliography}

\begin{thebibliography}{10}

\bibitem{PDF4LHCWorkingGroup:2022cjn}
R.~D. Ball {\em et~al.}, ``{The PDF4LHC21 combination of global PDF fits for
  the LHC Run III},'' {\em J. Phys. G}, vol.~49, no.~8, p.~080501, 2022.

\bibitem{Hou:2019efy}
T.-J. Hou {\em et~al.}, ``{New CTEQ global analysis of quantum chromodynamics
  with high-precision data from the LHC},'' {\em Phys. Rev. D}, vol.~103,
  no.~1, p.~014013, 2021.

\bibitem{Bailey:2020ooq}
S.~Bailey, T.~Cridge, L.~A. Harland-Lang, A.~D. Martin, and R.~S. Thorne,
  ``{Parton distributions from LHC, HERA, Tevatron and fixed target data:
  MSHT20 PDFs},'' {\em Eur. Phys. J. C}, vol.~81, no.~4, p.~341, 2021.

\bibitem{NNPDF:2021njg}
R.~D. Ball {\em et~al.}, ``{The path to proton structure at 1\% accuracy},''
  {\em Eur. Phys. J. C}, vol.~82, no.~5, p.~428, 2022.

\bibitem{McGowan:2022nag}
J.~McGowan, T.~Cridge, L.~A. Harland-Lang, and R.~S. Thorne, ``{Approximate
  N$^{3}$LO Parton Distribution Functions with Theoretical Uncertainties:
  MSHT20aN$^3$LO PDFs},'' 7 2022.

\bibitem{Sato:2019yez}
N.~Sato, C.~Andres, J.~J. Ethier, and W.~Melnitchouk, ``{Strange quark
  suppression from a simultaneous Monte Carlo analysis of parton distributions
  and fragmentation functions},'' {\em Phys. Rev. D}, vol.~101, no.~7,
  p.~074020, 2020.

\bibitem{Abdolmaleki:2021yjf}
H.~Abdolmaleki, M.~Soleymaninia, H.~Khanpour, S.~Amoroso, F.~Giuli, A.~Glazov,
  A.~Luszczak, F.~Olness, and O.~Zenaiev, ``{QCD analysis of pion fragmentation
  functions in the xFitter framework},'' {\em Phys. Rev. D}, vol.~104, no.~5,
  p.~056019, 2021.

\bibitem{Moffat:2021dji}
E.~Moffat, W.~Melnitchouk, T.~C. Rogers, and N.~Sato, ``{Simultaneous
  Monte~Carlo analysis of parton densities and fragmentation functions},'' {\em
  Phys. Rev. D}, vol.~104, no.~1, p.~016015, 2021.

\bibitem{Khalek:2021gxf}
R.~A. Khalek, V.~Bertone, and E.~R. Nocera, ``{Determination of unpolarized
  pion fragmentation functions using semi-inclusive deep-inelastic-scattering
  data},'' {\em Phys. Rev. D}, vol.~104, no.~3, p.~034007, 2021.

\bibitem{Borsa:2022vvp}
I.~Borsa, D.~de~Florian, R.~Sassot, M.~Stratmann, and W.~Vogelsang, ``{Towards
  a Global QCD Analysis of Fragmentation Functions at Next-to-Next-to-Leading
  Order Accuracy},'' {\em Phys. Rev. Lett.}, vol.~129, no.~1, p.~012002, 2022.

\bibitem{Khalek:2022vgy}
R.~A. Khalek, V.~Bertone, A.~Khoudli, and E.~R. Nocera, ``{Pion and kaon
  fragmentation functions at next-to-next-to-leading order},'' 4 2022.

\bibitem{Collins:1984kg}
J.~C. Collins, D.~E. Soper, and G.~F. Sterman, ``{Transverse Momentum
  Distribution in Drell-Yan Pair and W and Z Boson Production},'' {\em Nucl.
  Phys. B}, vol.~250, pp.~199--224, 1985.

\bibitem{Collins:2011zzd}
J.~Collins, {\em {Foundations of perturbative QCD}}, vol.~32.
\newblock Cambridge University Press, 11 2013.

\bibitem{Bacchetta:2017gcc}
A.~Bacchetta, F.~Delcarro, C.~Pisano, M.~Radici, and A.~Signori, ``{Extraction
  of partonic transverse momentum distributions from semi-inclusive
  deep-inelastic scattering, Drell-Yan and Z-boson production},'' {\em JHEP},
  vol.~06, p.~081, 2017.
\newblock [Erratum: JHEP 06, 051 (2019)].

\bibitem{Bacchetta:2019sam}
A.~Bacchetta, V.~Bertone, C.~Bissolotti, G.~Bozzi, F.~Delcarro, F.~Piacenza,
  and M.~Radici, ``{Transverse-momentum-dependent parton distributions up to
  N$^{3}$LL from Drell-Yan data},'' {\em JHEP}, vol.~07, p.~117, 2020.

\bibitem{Scimemi:2019cmh}
I.~Scimemi and A.~Vladimirov, ``{Non-perturbative structure of semi-inclusive
  deep-inelastic and Drell-Yan scattering at small transverse momentum},'' {\em
  JHEP}, vol.~06, p.~137, 2020.

\bibitem{Bacchetta:2022awv}
A.~Bacchetta, V.~Bertone, C.~Bissolotti, G.~Bozzi, M.~Cerutti, F.~Piacenza,
  M.~Radici, and A.~Signori, ``{Unpolarized Transverse Momentum Distributions
  from a global fit of Drell-Yan and Semi-Inclusive Deep-Inelastic Scattering
  data},'' 6 2022.

\bibitem{Boglione:2022nzq}
M.~Boglione, J.~O. Gonzalez-Hernandez, and A.~Simonelli, ``{Transverse Momentum
  Dependent Fragmentation Functions from recent BELLE data},'' 6 2022.

\bibitem{Ebert:2022cku}
M.~A. Ebert, J.~K.~L. Michel, I.~W. Stewart, and Z.~Sun, ``{Disentangling Long
  and Short Distances in Momentum-Space TMDs},'' 1 2022.

\bibitem{Ji:1998pc}
X.-D. Ji, ``{Off forward parton distributions},'' {\em J. Phys. G}, vol.~24,
  pp.~1181--1205, 1998.

\bibitem{Polyakov:2002yz}
M.~V. Polyakov, ``{Generalized parton distributions and strong forces inside
  nucleons and nuclei},'' {\em Phys. Lett. B}, vol.~555, pp.~57--62, 2003.

\bibitem{Burkardt:2002hr}
M.~Burkardt, ``{Impact parameter space interpretation for generalized parton
  distributions},'' {\em Int. J. Mod. Phys. A}, vol.~18, pp.~173--208, 2003.

\bibitem{Diehl:2002he}
M.~Diehl, ``{Generalized parton distributions in impact parameter space},''
  {\em Eur. Phys. J. C}, vol.~25, pp.~223--232, 2002.
\newblock [Erratum: Eur.Phys.J.C 31, 277--278 (2003)].

\bibitem{Ji:1996ek}
X.-D. Ji, ``{Gauge-Invariant Decomposition of Nucleon Spin},'' {\em Phys. Rev.
  Lett.}, vol.~78, pp.~610--613, 1997.

\bibitem{Kumericki:2015lhb}
K.~Kumeri\v{c}ki and D.~M\"uller, ``{Description and interpretation of DVCS
  measurements},'' {\em EPJ Web Conf.}, vol.~112, p.~01012, 2016.

\bibitem{Dutrieux:2021wll}
H.~Dutrieux, H.~Dutrieux, O.~Grocholski, O.~Grocholski, H.~Moutarde,
  H.~Moutarde, P.~Sznajder, and P.~Sznajder, ``{Artificial neural network
  modelling of generalised parton distributions},'' {\em Eur. Phys. J. C},
  vol.~82, no.~3, p.~252, 2022.
\newblock [Erratum: Eur.Phys.J.C 82, 389 (2022)].

\bibitem{Guo:2022upw}
Y.~Guo, X.~Ji, and K.~Shiells, ``{Generalized parton distributions through
  universal moment parameterization: zero skewness case},'' 7 2022.

\bibitem{Anderle:2021wcy}
D.~P. Anderle {\em et~al.}, ``{Electron-ion collider in China},'' {\em Front.
  Phys. (Beijing)}, vol.~16, no.~6, p.~64701, 2021.

\bibitem{AbdulKhalek:2021gbh}
R.~Abdul~Khalek {\em et~al.}, ``{Science Requirements and Detector Concepts for
  the Electron-Ion Collider: EIC Yellow Report},'' 3 2021.

\bibitem{Ji:2013dva}
X.~Ji, ``{Parton Physics on a Euclidean Lattice},'' {\em Phys. Rev. Lett.},
  vol.~110, p.~262002, 2013.

\bibitem{Radyushkin:2017cyf}
A.~V. Radyushkin, ``{Quasi-parton distribution functions, momentum
  distributions, and pseudo-parton distribution functions},'' {\em Phys. Rev.
  D}, vol.~96, no.~3, p.~034025, 2017.

\bibitem{Belitsky:2003nz}
A.~V. Belitsky, X.-d. Ji, and F.~Yuan, ``{Quark imaging in the proton via
  quantum phase space distributions},'' {\em Phys. Rev. D}, vol.~69, p.~074014,
  2004.

\bibitem{Ji:2003ak}
X.-d. Ji, ``{Viewing the proton through 'color' filters},'' {\em Phys. Rev.
  Lett.}, vol.~91, p.~062001, 2003.

\bibitem{Meissner:2008ay}
S.~Meissner, A.~Metz, M.~Schlegel, and K.~Goeke, ``{Generalized parton
  correlation functions for a spin-0 hadron},'' {\em JHEP}, vol.~08, p.~038,
  2008.

\bibitem{Meissner:2009ww}
S.~Meissner, A.~Metz, and M.~Schlegel, ``{Generalized parton correlation
  functions for a spin-1/2 hadron},'' {\em JHEP}, vol.~08, p.~056, 2009.

\bibitem{Lorce:2011kd}
C.~Lorce and B.~Pasquini, ``{Quark Wigner Distributions and Orbital Angular
  Momentum},'' {\em Phys. Rev. D}, vol.~84, p.~014015, 2011.

\bibitem{Lorce:2011ni}
C.~Lorce, B.~Pasquini, X.~Xiong, and F.~Yuan, ``{The quark orbital angular
  momentum from Wigner distributions and light-cone wave functions},'' {\em
  Phys. Rev. D}, vol.~85, p.~114006, 2012.

\bibitem{Lorce:2013pza}
C.~Lorc\'e and B.~Pasquini, ``{Structure analysis of the generalized correlator
  of quark and gluon for a spin-1/2 target},'' {\em JHEP}, vol.~09, p.~138,
  2013.

\bibitem{Lorce:2011dv}
C.~Lorce, B.~Pasquini, and M.~Vanderhaeghen, ``{Unified framework for
  generalized and transverse-momentum dependent parton distributions within a
  3Q light-cone picture of the nucleon},'' {\em JHEP}, vol.~05, p.~041, 2011.

\bibitem{Kanazawa:2014nha}
K.~Kanazawa, C.~Lorc\'e, A.~Metz, B.~Pasquini, and M.~Schlegel, ``{Twist-2
  generalized transverse-momentum dependent parton distributions and the
  spin/orbital structure of the nucleon},'' {\em Phys. Rev. D}, vol.~90, no.~1,
  p.~014028, 2014.

\bibitem{Lorce:2015sqe}
C.~Lorc\'e and B.~Pasquini, ``{Multipole decomposition of the nucleon
  transverse phase space},'' {\em Phys. Rev. D}, vol.~93, no.~3, p.~034040,
  2016.

\bibitem{Hagiwara:2016kam}
Y.~Hagiwara, Y.~Hatta, and T.~Ueda, ``{Wigner, Husimi, and generalized
  transverse momentum dependent distributions in the color glass condensate},''
  {\em Phys. Rev. D}, vol.~94, no.~9, p.~094036, 2016.

\bibitem{Hagiwara:2017fye}
Y.~Hagiwara, Y.~Hatta, R.~Pasechnik, M.~Tasevsky, and O.~Teryaev, ``{Accessing
  the gluon Wigner distribution in ultraperipheral $pA$ collisions},'' {\em
  Phys. Rev. D}, vol.~96, no.~3, p.~034009, 2017.

\bibitem{More:2017zqq}
J.~More, A.~Mukherjee, and S.~Nair, ``{Quark Wigner Distributions Using
  Light-Front Wave Functions},'' {\em Phys. Rev. D}, vol.~95, no.~7, p.~074039,
  2017.

\bibitem{More:2017zqp}
J.~More, A.~Mukherjee, and S.~Nair, ``{Wigner Distributions For Gluons},'' {\em
  Eur. Phys. J. C}, vol.~78, no.~5, p.~389, 2018.

\bibitem{Bhattacharya:2018zxi}
S.~Bhattacharya, C.~Cocuzza, and A.~Metz, ``{Generalized quasi parton
  distributions in a diquark spectator model},'' {\em Phys. Lett. B}, vol.~788,
  pp.~453--463, 2019.

\bibitem{Mantysaari:2019csc}
H.~M\"antysaari, N.~Mueller, and B.~Schenke, ``{Diffractive Dijet Production
  and Wigner Distributions from the Color Glass Condensate},'' {\em Phys. Rev.
  D}, vol.~99, no.~7, p.~074004, 2019.

\bibitem{Kaur:2019kpi}
N.~Kaur and H.~Dahiya, ``{Quark Wigner Distributions and GTMDs of Pion in the
  Light-Front Holographic Model},'' {\em Eur. Phys. J. A}, vol.~56, no.~6,
  p.~172, 2020.

\bibitem{Salazar:2019ncp}
F.~Salazar and B.~Schenke, ``{Diffractive dijet production in impact parameter
  dependent saturation models},'' {\em Phys. Rev. D}, vol.~100, no.~3,
  p.~034007, 2019.

\bibitem{Luo:2020yqj}
X.~Luo and H.~Sun, ``{T-odd generalized and quasi transverse momentum dependent
  parton distribution in a scalar spectator model},'' {\em Eur. Phys. J. C},
  vol.~80, no.~9, p.~828, 2020.

\bibitem{Zhang:2020ecj}
J.-L. Zhang, Z.-F. Cui, J.~Ping, and C.~D. Roberts, ``{Contact interaction
  analysis of pion GTMDs},'' {\em Eur. Phys. J. C}, vol.~81, no.~1, p.~6, 2021.

\bibitem{Boer:2021upt}
D.~Boer and C.~Setyadi, ``{GTMD model predictions for diffractive dijet
  production at EIC},'' {\em Phys. Rev. D}, vol.~104, no.~7, p.~074006, 2021.

\bibitem{Hatta:2016dxp}
Y.~Hatta, B.-W. Xiao, and F.~Yuan, ``{Probing the Small- x Gluon Tomography in
  Correlated Hard Diffractive Dijet Production in Deep Inelastic Scattering},''
  {\em Phys. Rev. Lett.}, vol.~116, no.~20, p.~202301, 2016.

\bibitem{Bhattacharya:2017bvs}
S.~Bhattacharya, A.~Metz, and J.~Zhou, ``{Generalized TMDs and the exclusive
  double Drell\textendash{}Yan process},'' {\em Phys. Lett. B}, vol.~771,
  pp.~396--400, 2017.
\newblock [Erratum: Phys.Lett.B 810, 135866 (2020)].

\bibitem{Bhattacharya:2018lgm}
S.~Bhattacharya, A.~Metz, V.~K. Ojha, J.-Y. Tsai, and J.~Zhou, ``{Exclusive
  double quarkonium production and generalized TMDs of gluons},'' 2 2018.

\bibitem{Echevarria:2016mrc}
M.~G. Echevarria, A.~Idilbi, K.~Kanazawa, C.~Lorc\'e, A.~Metz, B.~Pasquini, and
  M.~Schlegel, ``{Proper definition and evolution of generalized transverse
  momentum dependent distributions},'' {\em Phys. Lett. B}, vol.~759,
  pp.~336--341, 2016.

\bibitem{Bertone:2022frx}
V.~Bertone, H.~Dutrieux, C.~Mezrag, J.~M. Morgado, and H.~Moutarde,
  ``{Revisiting evolution equations for generalised parton distributions},'' 6
  2022.

\bibitem{Echevarria:2011epo}
M.~G. Echevarria, A.~Idilbi, and I.~Scimemi, ``{Factorization Theorem For
  Drell-Yan At Low $q_T$ And Transverse Momentum Distributions
  On-The-Light-Cone},'' {\em JHEP}, vol.~07, p.~002, 2012.

\bibitem{Collins:2003fm}
J.~C. Collins, ``{What exactly is a parton density?},'' {\em Acta Phys. Polon.
  B}, vol.~34, p.~3103, 2003.

\bibitem{Echevarria:2016scs}
M.~G. Echevarria, I.~Scimemi, and A.~Vladimirov, ``{Unpolarized Transverse
  Momentum Dependent Parton Distribution and Fragmentation Functions at
  next-to-next-to-leading order},'' {\em JHEP}, vol.~09, p.~004, 2016.

\bibitem{Ebert:2019okf}
M.~A. Ebert, I.~W. Stewart, and Y.~Zhao, ``{Towards Quasi-Transverse Momentum
  Dependent PDFs Computable on the Lattice},'' {\em JHEP}, vol.~09, p.~037,
  2019.

\bibitem{Idilbi:2010im}
A.~Idilbi and I.~Scimemi, ``{Singular and Regular Gauges in Soft Collinear
  Effective Theory: The Introduction of the New Wilson Line T},'' {\em Phys.
  Lett. B}, vol.~695, pp.~463--468, 2011.

\bibitem{Curci:1980uw}
G.~Curci, W.~Furmanski, and R.~Petronzio, ``{Evolution of Parton Densities
  Beyond Leading Order: The Nonsinglet Case},'' {\em Nucl. Phys. B}, vol.~175,
  pp.~27--92, 1980.

\bibitem{Ji:2002aa}
X.-d. Ji and F.~Yuan, ``{Parton distributions in light cone gauge: Where are
  the final state interactions?},'' {\em Phys. Lett. B}, vol.~543, pp.~66--72,
  2002.

\bibitem{Rodini:2021zcs}
S.~Rodini, {\em {Proton structure, an iridescent study: from parton
  distributions to the emergence of the proton mass}}.
\newblock PhD thesis, Universita' Di Pavia, Pavia U., 2021.

\bibitem{Collins:1981uk}
J.~C. Collins and D.~E. Soper, ``{Back-To-Back Jets in QCD},'' {\em Nucl. Phys.
  B}, vol.~193, p.~381, 1981.
\newblock [Erratum: Nucl.Phys.B 213, 545 (1983)].

\bibitem{Collins:1981va}
J.~C. Collins and D.~E. Soper, ``{Back-To-Back Jets: Fourier Transform from B
  to K-Transverse},'' {\em Nucl. Phys. B}, vol.~197, pp.~446--476, 1982.

\bibitem{Goeke:2006ef}
K.~Goeke, S.~Meissner, A.~Metz, and M.~Schlegel, ``{Checking the Burkardt sum
  rule for the Sivers function by model calculations},'' {\em Phys. Lett. B},
  vol.~637, pp.~241--244, 2006.

\bibitem{Bacchetta:2008xw}
A.~Bacchetta, D.~Boer, M.~Diehl, and P.~J. Mulders, ``{Matches and mismatches
  in the descriptions of semi-inclusive processes at low and high transverse
  momentum},'' {\em JHEP}, vol.~08, p.~023, 2008.

\bibitem{Collins:1984xc}
J.~C. Collins, {\em {Renormalization}: {An Introduction to Renormalization, The
  Renormalization Group, and the Operator Product Expansion}}, vol.~26 of {\em
  Cambridge Monographs on Mathematical Physics}.
\newblock Cambridge: Cambridge University Press, 1986.

\bibitem{Vladimirov:2014aja}
A.~A. Vladimirov, ``{TMD PDFs in the Laguerre polynomial basis},'' {\em JHEP},
  vol.~08, p.~089, 2014.

\bibitem{Echevarria:2015byo}
M.~G. Echevarria, I.~Scimemi, and A.~Vladimirov, ``{Universal transverse
  momentum dependent soft function at NNLO},'' {\em Phys. Rev. D}, vol.~93,
  no.~5, p.~054004, 2016.

\bibitem{Li:2016axz}
Y.~Li, D.~Neill, and H.~X. Zhu, ``{An exponential regulator for rapidity
  divergences},'' {\em Nucl. Phys. B}, vol.~960, p.~115193, 2020.

\bibitem{Deng:2022gzi}
Z.-F. Deng, W.~Wang, and J.~Zeng, ``{Transverse-momentum-dependent wave
  functions and Soft functions at one-loop in Large Momentum Effective
  Theory},'' 7 2022.

\bibitem{Li:2011zp}
Y.~Li, S.~Mantry, and F.~Petriello, ``{An Exclusive Soft Function for Drell-Yan
  at Next-to-Next-to-Leading Order},'' {\em Phys. Rev. D}, vol.~84, p.~094014,
  2011.

\bibitem{Li:2016ctv}
Y.~Li and H.~X. Zhu, ``{Bootstrapping Rapidity Anomalous Dimensions for
  Transverse-Momentum Resummation},'' {\em Phys. Rev. Lett.}, vol.~118, no.~2,
  p.~022004, 2017.

\bibitem{Ebert:2020yqt}
M.~A. Ebert, B.~Mistlberger, and G.~Vita, ``{Transverse momentum dependent PDFs
  at N$^3$LO},'' {\em JHEP}, vol.~09, p.~146, 2020.

\bibitem{Bertone:2021yyz}
V.~Bertone, H.~Dutrieux, C.~Mezrag, H.~Moutarde, and P.~Sznajder,
  ``{Deconvolution problem of deeply virtual Compton scattering},'' {\em Phys.
  Rev. D}, vol.~103, no.~11, p.~114019, 2021.

\bibitem{Diehl:2003ny}
M.~Diehl, ``{Generalized parton distributions},'' {\em Phys. Rept.}, vol.~388,
  pp.~41--277, 2003.

\bibitem{Kivel:2000rb}
N.~Kivel, M.~V. Polyakov, A.~Schafer, and O.~V. Teryaev, ``{On the
  Wandzura-Wilczek approximation for the twist - three DVCS amplitude},'' {\em
  Phys. Lett. B}, vol.~497, pp.~73--79, 2001.

\bibitem{Kivel:2000cn}
N.~Kivel and M.~V. Polyakov, ``{DVCS on the nucleon to the twist - three
  accuracy},'' {\em Nucl. Phys. B}, vol.~600, pp.~334--350, 2001.

\bibitem{Radyushkin:2000jy}
A.~V. Radyushkin and C.~Weiss, ``{DVCS amplitude with kinematical twist - three
  terms},'' {\em Phys. Lett. B}, vol.~493, pp.~332--340, 2000.

\bibitem{Altinoluk:2012nt}
T.~Altinoluk, B.~Pire, L.~Szymanowski, and S.~Wallon, ``{Resumming soft and
  collinear contributions in deeply virtual Compton scattering},'' {\em JHEP},
  vol.~10, p.~049, 2012.

\bibitem{Catani:1989ne}
S.~Catani and L.~Trentadue, ``{Resummation of the QCD Perturbative Series for
  Hard Processes},'' {\em Nucl. Phys. B}, vol.~327, pp.~323--352, 1989.

\bibitem{Catani:1990rp}
S.~Catani and L.~Trentadue, ``{Comment on QCD exponentiation at large x},''
  {\em Nucl. Phys. B}, vol.~353, pp.~183--186, 1991.

\bibitem{Collins:2017oxh}
J.~Collins and T.~C. Rogers, ``{Connecting Different TMD Factorization
  Formalisms in QCD},'' {\em Phys. Rev. D}, vol.~96, no.~5, p.~054011, 2017.

\bibitem{Bozzi:2005wk}
G.~Bozzi, S.~Catani, D.~de~Florian, and M.~Grazzini, ``{Transverse-momentum
  resummation and the spectrum of the Higgs boson at the LHC},'' {\em Nucl.
  Phys. B}, vol.~737, pp.~73--120, 2006.

\bibitem{Bizon:2018foh}
W.~Bizo\'n, X.~Chen, A.~Gehrmann-De~Ridder, T.~Gehrmann, N.~Glover, A.~Huss,
  P.~F. Monni, E.~Re, L.~Rottoli, and P.~Torrielli, ``{Fiducial distributions
  in Higgs and Drell-Yan production at N$^{3}$LL+NNLO},'' {\em JHEP}, vol.~12,
  p.~132, 2018.

\bibitem{Goloskokov:2005sd}
S.~V. Goloskokov and P.~Kroll, ``{Vector meson electroproduction at small
  Bjorken-x and generalized parton distributions},'' {\em Eur. Phys. J. C},
  vol.~42, pp.~281--301, 2005.

\bibitem{Goloskokov:2007nt}
S.~V. Goloskokov and P.~Kroll, ``{The Role of the quark and gluon GPDs in hard
  vector-meson electroproduction},'' {\em Eur. Phys. J. C}, vol.~53,
  pp.~367--384, 2008.

\bibitem{Goloskokov:2009ia}
S.~V. Goloskokov and P.~Kroll, ``{An Attempt to understand exclusive pi+
  electroproduction},'' {\em Eur. Phys. J. C}, vol.~65, pp.~137--151, 2010.

\bibitem{Berthou:2015oaw}
B.~Berthou {\em et~al.}, ``{PARTONS: PARtonic Tomography Of Nucleon Software}:
  {A computing framework for the phenomenology of Generalized Parton
  Distributions},'' {\em Eur. Phys. J. C}, vol.~78, no.~6, p.~478, 2018.

\bibitem{Bertone:2013vaa}
V.~Bertone, S.~Carrazza, and J.~Rojo, ``{APFEL: A PDF Evolution Library with
  QED corrections},'' {\em Comput. Phys. Commun.}, vol.~185, pp.~1647--1668,
  2014.

\bibitem{Bertone:2017gds}
V.~Bertone, ``{APFEL++: A new PDF evolution library in C++},'' {\em PoS},
  vol.~DIS2017, p.~201, 2018.

\bibitem{Aybat:2011zv}
S.~M. Aybat and T.~C. Rogers, ``{TMD Parton Distribution and Fragmentation
  Functions with QCD Evolution},'' {\em Phys. Rev. D}, vol.~83, p.~114042,
  2011.

\end{thebibliography}

\end{document}